\documentclass[10pt,prd,twocolumn,notitlepage,floatfix,shownopacs,preprintnumbers,nofootinbib,longbibliography]{revtex4-2}
\pdfoutput=1
\usepackage{physics} 
\usepackage{txfonts}
\usepackage[bbgreekl]{mathbbol}
\usepackage{graphicx}
\graphicspath{{figs/}}
\usepackage{url}
\usepackage{color}
\usepackage[colorlinks=true,breaklinks=true]{hyperref}
\hypersetup{allcolors=[rgb]{0.0 0.0 0.6},linkcolor=[rgb]{0.8 0.0 0.4}}   
\allowdisplaybreaks

\bibpunct{[}{]}{,}{n}{}{,}

\renewcommand{\pv}{\text{P.V.}}

\usepackage{mathtools} 
\usepackage{xparse} 
\makeatletter
\newcommand{\raisemath}[1]{\mathpalette{\raisem@th{#1}}} 
\newcommand{\raisem@th}[3]{\raisebox{#1}{$#2#3$}}
\makeatother
\NewDocumentCommand{\dbar}{O{0pt} O{0pt}}{
  \ensuremath{\mathrlap{\raisemath{-0.25pt}{\hspace*{2.9pt}{\mathchar'26\mkern-9mu}}}d}%
}

\newcommand{\bp}{{\bf p}}
\newcommand{\bx}{{\bf x}}
\newcommand{\bv}{{\bf v}}
\newcommand{\bk}{{\bf k}}

\newcommand{\bG}{{\bf\Gamma}}
\providecommand{\varw}{w}

\newcommand{\sH}{{\sf H}}
\newcommand{\sM}{{\sf M}}

\newcommand{\GF}{G_{\rm F}}
\newcommand{\we}{{\varw_E}}
\newcommand{\mD}{{\mathcal{D}}}
\newcommand{\mI}{{\mathcal{I}}}
\newcommand{\mCd}{{\mathcal{C}_{\mathcal{D}}}}
\newcommand{\mCw}{{\mathcal{C}_{\omega}}}

\definecolor{blue}{rgb}{0.0,0.0,1.0}

\definecolor{red}{rgb}{1.0,0.0,0.0}

\usepackage{tikz,xcolor,hyperref}
\definecolor{lime}{HTML}{A6CE39}
\DeclareRobustCommand{\orcidicon}{\hspace{-1mm}
	\begin{tikzpicture}
	\draw[lime, fill=lime] (0,0) 
	circle [radius=0.16] 
	node[white] {{\fontfamily{qag}\selectfont \tiny \,ID}};
	\draw[white, fill=white] (-0.0525,0.095) 
	circle [radius=0.007];
	\end{tikzpicture}
	\hspace{-3mm}
}
\foreach \x in {A, ..., Z}{\expandafter\xdef\csname orcid\x\endcsname{\noexpand\href{https://orcid.org/\csname orcidauthor\x\endcsname}
			{\noexpand\orcidicon}}
}

\begin{document}

\title{Sufficient and Necessary Conditions for Collective Neutrino Instability: Fast, Slow, and Mixed}

\author{Basudeb Dasgupta\orcidA{}}
\email{bdasgupta@theory.tifr.res.in}

\author{Dwaipayan Mukherjee\orcidB{}}
\email{dwaipayan.mukherjee@tifr.res.in}

\affiliation{Tata Institute of Fundamental Research, Homi Bhabha Road, Mumbai, 400005, India}

\date[First version:\,]{May 6, 2025; This version: November 25, 2025: Accepted: December 2, 2025}

\begin{abstract}Collective neutrino oscillations exhibit instabilities that induce appreciable flavor conversion, with crucial astrophysical implications. While the importance of initial phase-space distributions is well-established, a general instability criterion for distributions dependent on both energy and emission angle has been lacking. We identify and analyze the sufficient (and necessary) conditions for a generic collective neutrino flavor instability.
\end{abstract}

\preprint{TIFR/TH/25-12}

\maketitle

\tableofcontents

\section{Introduction}
\label{sec:intro}
Dense neutrino gases can exhibit collective flavor conversion due to forward scattering of neutrinos and antineutrinos off each other~\cite{Volpe:2023met,Johns:2025mlm}. Although ordinary flavor mixing is highly suppressed in these environments~\cite{Wolfenstein:1977ue, Mikheev:1987jp, Stodolsky:1986dx}, instabilities due to collective evolution can lead to appreciable flavor change~\cite{Pantaleone:1992xh, Pantaleone:1992eq}. A key challenge has been identifying the conditions under which such instabilities arise.

A widely held conjecture is that a {zero-crossing} in the distribution of flavor differences~\cite{Dasgupta:2009mg} determines the presence and nature of instabilities --- an idea supported by studies of specific examples in both slow\,\cite{Banerjee:2011fj} and fast~\cite{Capozzi:2017gqd, Abbar:2017pkh, Capozzi:2019lso,Yi:2019hrp}
collective oscillations. The necessary and sufficient criterion for fast instability~\cite{Morinaga:2021vmc}, and a necessary condition for the general case (i.e., including both slow and fast instability)~\cite{Dasgupta:2021gfs} are known, but the sufficient criteria for collective instability in the general case still remain unknown.

In this paper, we rigorously establish, analyze, and explain the conditions that are \emph{sufficient} (and necessary) for collective instability in general --- whether they are fast, slow, or a combination of the two. We find that zero-crossing alone is not sufficient for instability, in general, and we identify the additional conditions that must be met. We dissect the additional conditions to show why, in the case of fast instability, they are identically satisfied, whereas for slow or mixed instability, they typically are not. We then show how the additional conditions can be satisfied when the collective potential significantly exceeds the vacuum oscillation frequency. We examine a few special cases and representative examples to highlight the meaning and importance of our findings. We then compare our results with relevant previous work, highlighting key similarities and differences.

\section{Framework and Notation}
\label{sec:frmwrk}
At a location $\bx$ and time $t$,  the flavor content of neutrinos of momentum ${\bf p}$ is encoded in the Wigner distributions  or ``density matrices'' $\rho_{\bp}$ that evolve as 
\begin{equation}
  i\,(\partial_t+ \bv\cdot{\partial_{\bx}})\,\rho_{\bp} = [\sH_{\bp},\rho_{\bp}]\,,
    \label{eq:EOM}
\end{equation}
in natural units with $\hbar=c=1$. The effective Hamiltonian has contributions from the neutrino mass-mixing, as well as interactions with ordinary matter and other neutrinos,
\begin{equation}
    \sH_{\bp} = \frac{\sM^2}{2E} + \sH_{\bp}^{\rm M} + \sH_{\bp}^{\nu\nu}\,.
    \label{eq:hamiltonian}
\end{equation}
The collective term is $\sH^{\nu\nu}_\bp=\sqrt{2}\GF\,v_\alpha \int \dbar\bp'\,v'^\alpha\left(\rho_{\bp'}-\bar\rho_{\bp'}\right)$, where \mbox{$\dbar\bp={d^3\bp/(2\pi)^3}$} and $v^\alpha=(1,\bv)$ the neutrino four-velocity. The matter term {$\sH^{\rm M}_\bp$ is diagonal in the flavor basis with elements $\sqrt{2}\GF\,{(s_\ell)}_\alpha\int \dbar\bp^\alpha_\ell\left(f_{\ell,\bp}-\bar f_{\ell,\bp}\right)$} for the $\ell^{\rm th}$ charged lepton having a phase space distribution $f_{\ell,\bp}$ and a four-velocity {$s_\ell^\alpha$}. The mass-mixing term does not depend on $\bv$, and the refractive term does not depend on $E$, but only on $\bv$. The equation for antineutrinos is the same, except for a minus sign for the ${\sM^2}/{(2E)}$ term. Thus, antineutrinos of energy $+E$ can be treated as neutrinos of negative energy $-E$. All dimensional quantities are hereafter in units of a nominal collective oscillation frequency scale $\mu\equiv2\sqrt{2}G_{\rm F}(n_{\nu_e}+\,n_{\bar\nu_e})$. 

We are interested in the following question: what are the conditions under which the flavor distributions begin to change exponentially in time? We consider Eq.\,(\ref{eq:EOM}) in the limit of vanishing flavor-mixing, linearize it in the off-diagonal element $S_{\hspace{-0.5mm}\bp}$ for any two flavors, and solve for its Fourier modes $\sim e^{-i(\omega t- \bk \cdot \bx)}$, following Ref.\,\cite{Dasgupta:2021gfs}; see refs.\,\cite{Banerjee:2011fj, Izaguirre:2016gsx, Airen:2018nvp} for prior work. We find nontrivial solutions require that the frequency $\omega$ must be related to the (shifted) wavevector $\bk$ via
\begin{equation}
\text{det}\,\Pi(\omega,\bk)=0\,,
\end{equation}
where
\begin{equation}
    \Pi^{\alpha\beta}(\omega,\bk) = \eta^{\alpha\beta} + \int d\bG\; g_{\bG} \dfrac{v^{\alpha}v^{\beta}}{\omega -\bk\cdot\bv - \we}\,,
\label{eq:Pidef}
\end{equation}
with $\eta={\rm diag}(+,-,-,-)$ being the metric tensor, $\we\equiv\Delta m^2/(2E)$ the vacuum oscillation frequency, $\bG$ the coordinate in the joint momentum space of neutrinos and antineutrinos, and $g_\bG$ the flavor-difference distribution.

We adopt the convention of Ref.\,\cite{Dasgupta:2021gfs}, where $\Delta m^2 = m_1^2-m_2^2$, which is positive for inverted mass ordering $(m_1>m_2)$ and negative for normal mass ordering $(m_1<m_2)$. We will, for definiteness, assume the inverted ordering. Thus, $\we$ is positive for neutrinos and negative for antineutrinos.

The three-dimensional momentum-space is written using a signed inverse-radius-like coordinate $\we\in\mathbb{R}^1$, and by the angular coordinates $\bv \in \mathbb{S}^2$, such that 
\begin{equation}
    \int_{\mathbb{\Gamma}} d{\bG} \left[\cdots\right]=  \int_{\mathbb{R}^1} \dfrac{(\Delta m^2)^3}{64\pi^3\we^4}d\we \int_{\mathbb{S}^2} d\bv \left[\cdots\right]\,.
\label{eq:gammaspace}
\end{equation}

Neutrinos are initially in almost perfect flavor eigenstates; thus, their density matrices are diagonal. The distribution $g_\bG$ packages the initial phase space distributions of the two relevant flavors of neutrinos and antineutrinos, as
\begin{equation}
g_\bG = g({\we,\bv})= \begin{cases} f_{\nu_e}(\bp) - f_{\nu_\mu}(\bp)  \quad \text{for}\,\we>0\\ f_{\bar{\nu}_\mu}(\bp) -f_{\bar{\nu}_e}(\bp)  \quad \text{for}\,\we<0\,,\end{cases}
\label{eq:ggamma}
\end{equation}
with $\bp=|\Delta m^2/(2\we)|\bv$. For ultra-relativistic particles with finite average energies, the distributions $f_{\nu,\bar{\nu}}$ vanish at $E\to0$ and $E\to\pm\infty$. We assume that the fall-offs of $g_\bG$ at $\we\to\pm \infty$ and $\we\to0$, are rapid enough that all relevant integrals stay bounded. We also assume that $g_\bG$ is suitably well-behaved, i.e., without corners (undefined gradients) or flat crossings (vanishing gradient at zero-crossings).

\section{Instability Conditions}
\label{sec:suffcon}

We may interpret the $\text{det}\,\Pi=0$ condition to require nontrivial eigenvectors $a^\alpha$, such that 
\begin{equation}
a_\alpha^* \Pi^{\alpha\beta} a_\beta=0\,.
\end{equation}
For definiteness, consider time-like eigenvectors of the shape $a^\alpha = (1,{\bf a})$, with $1-|{\bf a}|^2>0$, that require 
\begin{equation}
    1\;+\;\int d\bG\, \dfrac{|1-{\bf a}\cdot{\bf v}|^2}{1-|{\bf a}|^2}\,\frac{g_{\bG}}{\omega -\bk\cdot\bv - \we}=0\,.
\label{eq:scDR}    
\end{equation}
Up to a positive function which can be absorbed into a redefinition of $g_{\bG}$, this is identical to the dispersion relation
\begin{equation}
    \mD(\omega, \bk) \equiv 1\;+\;\int d\bG \dfrac{g_{\bG}}{\omega -\bk\cdot\bv - \we}=0\,.
    \label{eq:disprel}
\end{equation}
Thus, to study stability for an unspecified $g_\bG$, it suffices to analyze the behavior of this dispersion relation. One asks -- when does one obtain a solution $\omega$ such that $\Im\omega>0$ for some $\bk\in\mathbb{R}^3$. Evidently, such a solution represents an exponentially growing linear instability. 

To begin with, one must note that Eq.\,(\ref{eq:disprel}) is not well-defined when the denominator of the integrand therein vanishes. For ease of discussion, let's define the function
\begin{equation}
h_\bG\equiv h(\we,\bv)= \bk\cdot\bv + \we\,.
\end{equation}
The integrand is singular at $\bG$ wherever $h_\bG=\omega$. On this ``level-set'' of $h_\bG$, the integral needs to be suitably regulated. Consider the dispersion relation
\begin{equation}
    \mD (\omega+i0^+, \bk) \equiv 1 + \mI(\omega,\bk)\,.
\end{equation}
At some fixed $\bk$, we can write
\begin{equation}
\mI(\omega)=\mI_{\rm PV}(\omega)+i\,\mI_{\delta}(\omega)\,,
\end{equation}
where, using the Sokhotski-Plemelj formula, one has\,\cite{garfken67:math}
\begin{align}
\mI_{\rm PV}(\omega)&=\pv\int  d\bG \dfrac{g_{\bG}}{\omega -h_\bG}\\
    &=\lim_{\epsilon\to0^+}\int_{|h_\bG-\omega|>\epsilon{\textstyle{\phantom{A^B}}}}\hspace{-7mm}d\bG \dfrac{g_{\bG}}{\omega -h_\bG}\,,\label{eq:Ipv2}\\
\mI_{\delta}(\omega)&=-\pi\int d\bG \, \delta(\omega - h_\bG) \, g_\bG\label{eq:Id1}\\ 
&=- \pi \int_{S_{\hspace{-0.4mm}\omega}} \frac{g_\bG}{|\nabla_\bG h_\bG|} \, d{S_{\hspace{-0.4mm}\omega}} \label{eq:Id2}\,.
\end{align}
The last integral is constrained on the level-surface ${S_{\hspace{-0.4mm}\omega}}$, on which $h_\bG=\omega$. The gradient operator in $\mathbb{\Gamma}$  is resolved as
\begin{equation}
\nabla_\bG=-\text{sgn}(\we)\,\we^2\,{\bv}\,\partial_\we + \we\nabla_{\bv}\,,
\end{equation}
which has components along $\bv$ and along $\nabla_\bv$ (which are orthogonal on $\mathbb{S}^2$). For the function $h_\bG$, one finds
\begin{equation}
\nabla_\bG h_\bG = - \text{sgn}(\we)\,\we^2\,{\bf v} + \we\left({\bk}-(\bk\cdot\bv)\bv\right)\,,
\end{equation}
where we used $\nabla_\bv(\bk\cdot\bv)=\left({\bk}-(\bk\cdot\bv)\bv\right)\equiv\bk_\perp$, i.e., the part of the vector $\bk$ that is perpendicular to $\bv$.

\subsection{Statement of Sufficiency Condition} 
\label{sec:prop_suff_cond}

{The dispersion relation $\mD (\omega,\bk)=0$, admits a complex root $\omega$ with $\Im\omega>0$ at some wave-vector $\bk$, for a time-like eigenvector  and the inverted mass ordering, \emph{if}
\begin{enumerate}
\item[{\bf \textit{1a:}}] the distribution function $g_{\bG}$ has a zero-crossing at \emph{some} point $\bG_0 = (\we_{_0},\bv_0)$, i.e., $g(\we_{_0},\bv_0)=0$ and $g_\bG$ takes both signs in the neighborhood,
\item[{\bf \textit{1b:}}] where the gradients of $g_\bG$ and $h_\bG=\bk\cdot\bv+\we$ have a positive dot-product, i.e., $(\nabla_\bG g_\bG)_0\cdot(\nabla_\bG h_\bG)_0>0$,~\text{and}
\item[{\bf \textit{2a:}}]  the principal value {$\mI_{\rm PV}(\omega_0)<-1$} at a frequency $\omega_0=h_{\bG_{0}}$, 
where $\Im\mI(\omega_0)=0$,
\item[{\bf \textit{2b:}}] while $\mI_{\rm PV}(\omega_i)>-1$, with $i=1,\,2,\,\ldots$, for 
any other frequencies $\omega_i$ where $\Im\mI(\omega_i)=0$.
\end{enumerate}}

The conditions in the above proposition may not be illuminating at this stage, but it is clear that they are more restrictive than requiring a zero-crossing of $g_\bG$. Note that Conditions 1a and 1b are local, whereas 2a and 2b involve integrals over $g_\bG$ and are global. In addition, they depend on the eigenvector $a^\alpha$ and wavevector $\bk$. Their physical content -- especially of Conditions 1b, 2a, and 2b -- will become clearer in special limits, such as fast and slow instabilities, where they simplify. One may also wonder, which of the possibly uncountably many crossing points of $g_\bG$ is the point $\bG_0$, and how does one find it? We will address this question as well. First, we prove the above proposition.

\medskip

\subsection{Proof of Sufficiency Condition} 
\label{sec:proof_suff_cond}
For $\omega$ in the upper half plane and at a fixed $\bk$, $\mD(\omega)$ is an analytic function of $\omega$.
It can have an imaginary part whenever the $\delta$-function in Eq.\,(\ref{eq:Id1}) is satisfied, i.e., $\omega=h_{\bG}$ for $\omega\in\mathbb{R}$. For  $\omega\in\mathbb{R}$ on either side of $\omega_0=\bk\cdot\bv_0 + \we_{_0}$, with $(\we_{_0},\bv_0)$ being a zero-crossing of $g_\bG$, we find $\Im\mD(\omega)$ flips its sign. Thus, as $\omega$ traces a closed contour encompassing the entire upper half plane, $\mD$ traces a closed curve on the complex $\mD$ plane, crossing the real axis at least twice (once as $\omega$ crosses $\omega_0$, and at $|\omega|\to\infty$ where $\mD\to1$). If this curve traced by $\mD$ encircles the origin (which we show is the case, under the conditions provided), its argument has changed by at least $2\pi$ and Cauchy's argument principle then requires at least one root of $\mD(\omega,\bk)$ inside the contour traced by $\omega$ (which was the upper half plane), thus proving what was to be proved. We now explicitly carry out these steps.

 \subsubsection{Sign Flip and Localization}
\label{sec:localize} 

Condition\,1a provides that there is a crossing of $g_\bG$ at $\bG_0$. For some fixed $\bk$, we can find a frequency $\omega_0=h_{\bG_0}$. We will now analyze how $\Im\mI(\omega)$ behaves around $\omega_0$. To that end, consider two values $\omega_\pm \equiv \omega_0 \pm \delta$, shifted by some small positive $\delta$ to the right or left of $\omega_0$. To linear order in $\delta$, an ansatz for a coordinate on $S_{\hspace{-0.5mm}\omega_\pm}$ is given by
\begin{equation}
\bG_\pm = \bG_0 \pm \frac{(\nabla_\bG h_\bG)_0}{|\nabla_\bG h_\bG|^2_0}\,\delta\,.
\label{eq:coordguess}
\end{equation}
It is easily seen that
\begin{equation}
h_{\bG_\pm} = h_{\bG_0} + (\nabla_\bG h_\bG)_0 \cdot (\bG_\pm - \bG_0) = \omega_0 \pm \delta\,,
\end{equation}
confirming that $\bG_\pm$ lie on $S_{\hspace{-0.5mm}\omega_\pm}$. Similarly, we can write
\begin{equation}
g_{\bG_\pm} = (\nabla_\bG g_\bG)_0 \cdot (\bG_\pm - \bG_0) = \pm \delta \, \frac{(\nabla_\bG g_\bG)_0 \cdot (\nabla_\bG h_\bG)_0}{|\nabla_\bG h_\bG|_0^2},
\end{equation}
to linear order in $\delta$, where we used that $g_\bG$ vanishes at $\bG_0$, i.e., $g_{\bG_0}=0$, as per Condition 1a.

Let $ \Omega_{\pm} \subset {S}_{\hspace{-0.5mm}\omega_\pm}$ be sufficiently small $\epsilon$-radius discs around $\bG_\pm$ ($\epsilon\ll\delta$), approximating the surfaces ``near" them, and ${\Omega}'_{\pm} = {S}_{\hspace{-0.5mm}\omega_\pm} \setminus \Omega_{\pm}$ be the complement ``far'' part of the surfaces. The near-contributions to the integrals are simply
\begin{equation}
\int_{\Omega_\pm} \frac{g_\bG}{|\nabla_\bG h_\bG|} \, dS_{\hspace{-0.5mm}\omega_\pm} = \pm \pi \epsilon^2 \delta \, \frac{(\nabla_\bG g_\bG)_0 \cdot (\nabla_\bG h_\bG)_0}{|\nabla_\bG h_\bG|_0^3},
\label{eq:nearcon}
\end{equation}
which give via Eq.\,(\ref{eq:Id2}),
\begin{equation}
{\Im}\mI(\omega_\pm) = \mp C  \epsilon^2 \delta + \text{far-contribution from }\,{\Omega}'_\pm\,,
\label{eq:Cd3far}
\end{equation}
with $\text{sgn}(C)=\text{sgn}\big[(\nabla_\bG g_\bG)_0 \cdot (\nabla_\bG h_\bG)_0\big]$. Note that generically the two gradients are not orthogonal (see Sec.\,\ref{sec:positivity} for more details), and the near-contribution to $\Im\mI(\omega)$ changes sign across $\omega_0$. 
$\Im\mI(\omega)$, as seen in Eq.\,(\ref{eq:Id2}), gets contributions from the full two-dimensional level-surface. The level-surface, however, intersects the two-dimensional crossing-surface of $g_\bG$ along a one-dimensional curve $\gamma_{\rm int}$ (or a set of such curves). The above analysis applies at any point on $\gamma_{\rm int}$. At a generic such point, the integral picks up substantial far contributions and $\Im\mI$ does not vanish. Given that $\mD$  is analytic, $\Im\mD$ can change sign only a finite number of times, at frequencies denoted by $\omega_i$. In the following, we focus on such discrete instances of $\omega_i$.

Condition 2a provides that there is \emph{some} $\omega_0$ where the full $\Im\mI(\omega_0)$ vanishes (not just the near contribution). In Eq.\,(\ref{eq:Cd3far}), the vanishing of $\Im\mI$ dictates that the far contribution must also vanish at $\omega_0$.  
At the shifted points $\omega_\pm$, the near contribution must now dominate by continuity of $\Im\mI(\omega)$.\footnote{We thank Damiano Fiorillo and Georg Raffelt for discussions that helped to clarify this point significantly.}
Seen another way, the gradients of $h_\bG$ and $g_\bG$ are positively aligned at the crossing point $\bG_0$, according to Condition\,1b. Here, $\Im\mI(\omega_0=h_{\bG_0})$ vanishes. Shifting infinitesimally away to $\omega_\pm$ generates the near contributions $\mp C \epsilon^2 \delta$, but now the far contributions must stay small by continuity of $\Im\mI$. Thus, $\Im\mI(\omega)$ must go from positive to negative as $\omega$ goes from $\omega_-$ to $\omega_+$. Where exactly is this crossing point $\bG_0$ corresponding to $\omega_0$, and how does one find it, one may ask? An $\omega_0$ across which $\Im\mI$ changes its sign, necessarily corresponds to \emph{some} crossing point of $g_\bG$; this is guaranteed by the intermediate value theorem applied to Eq.\,(\ref{eq:Id2}). However, without solving the dispersion relation, the location $\bG_0$ and its corresponding $\omega_0$ are not predicted.

\begin{figure*}
    \includegraphics[width=0.9\linewidth]{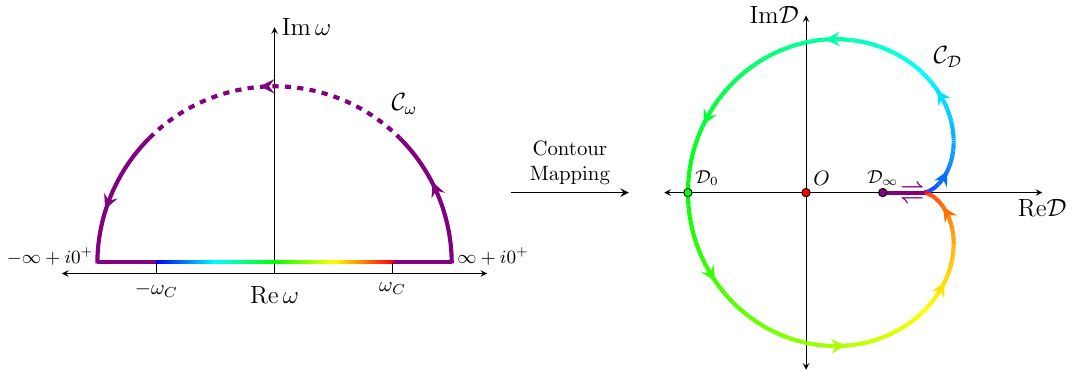}
    \caption{Sketch of contour mapping from $\mCw$ to $\mCd$. The color gradient from blue to red indicates the direction of the contour, with only a finite range of $\omega \in (-\omega_C,+\omega_C)$ allowing a nonzero $\Im\mD$. Crossing points of $\mCd$ are denoted by $\mD_i$, corresponding to $\omega_i$ where $\Im\mI$ changes sign. The contour $\mCd$ begins and terminates at $\mD_\infty=1$, as the semicircular arc $|\omega|\to\infty$ on $\mCw$ gets mapped to that point indicated by the purple dot. Note the flat region near it, where for large $|\omega|>\omega_C$, the dispersion $\mD$ is a real function.}
    \label{fig:contmap}
\end{figure*}

\begin{figure*}
    \includegraphics[width=0.9\linewidth]{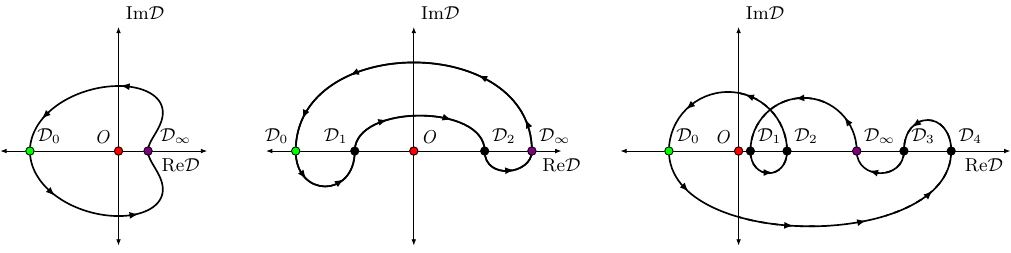}
    \caption{Sketch of $\mCd$ contours for singly and multiply crossed $g_\bG$, showing the relevance of Conditions 2a and 2b. For a singly crossed $g_\bG$, if Condition 2a is satisfied, i.e., the principal value corresponding to $\mI_{\rm PV}(\omega_0)<-1$, encirclement of origin is guaranteed (left panel). With multiple crossings, there may be more than one crossing to the left of the origin. If Condition 2b, viz. $\mI_{\rm PV}(\omega_{i\neq0})>-1$, is not satisfied; there may not be an encirclement of the origin (middle panel); if 2b is satisfied as well, encirclement cannot be avoided (e.g., right panel).
    \label{fig:oricirc}}
\end{figure*}

\subsubsection{Contour Mapping, Winding, and Zeros}
\label{sec:contour}

$\mD(\omega, \bk)$ is analytic in $\omega$ and has no poles for $\Im\omega > 0$. Thus, for fixed $\bk$, as $\omega$ traces a closed contour $\mCw$ in the complex-$\omega$ plane, $\mD(\omega + i0^+)$ traces a closed contour $\mCd$ in the complex-$\mD$ plane. Let $\mCw$ run just above the real axis from $-\infty+i0^+$ to $+\infty+i0^+$, closing via a large semicircle at infinity in the upper half-plane and returning to $-\infty+i0^+$; see left panel of Fig.\,\ref{fig:contmap}, with corresponding $\mCd$ on its right.

As $\omega$ moves rightward just above the real axis, it may encounter points $\omega_i$ where $\Im\mI(\omega_i)$ changes sign. At these points, $\mCd$ crosses the real axis. As we just showed, crossings from the upper to the lower half-plane correspond to Condition 1b being satisfied; the opposite indicates a violation. We label the crossing values of $\mD$ as $\mD_i$, with $\mD_0$ being the left-most crossing, consistent with Condition 2b. At large $|\omega|$, the principal value term $\mI_{\rm PV} \sim 1/|\omega| \to 0$, so $\Re\mD \to 1$, and $\mCd$ crosses the real axis at $\mD_\infty = 1$.

If $g_\bG$ has a single crossing, the contour $\mCd$ intersects the real axis at $\mD_0 = 1 + \mI_{\rm PV}(\omega_0)$, which lies to the left of the origin by Condition 2a, requiring $\mI_{\rm PV}(\omega_0) < -1$. The contour also crosses at $\mD_\infty = 1$ to the right of the origin, and nowhere else, thus necessarily encircling the origin, as shown in the left panel of Fig.\,\ref{fig:oricirc}. If there are multiple crossings, Condition 2a alone does not guarantee encirclement (see the middle panel of Fig.\,\ref{fig:oricirc} for an example). However, Condition 2b ensures $\mI_{\rm PV}(\omega_i) > -1$ for all other crossings, so that $\mD_i > 0$. With this, encirclement of the origin is again guaranteed; see, e.g., the right panel of Fig.\,\ref{fig:oricirc}.

Cauchy's argument principle states that for any function $F(z)$ meromorphic inside and on a closed contour, with no zeros or poles on the contour~\cite{garfken67:math},
\begin{equation}
    \ointctrclockwise \,dz \dfrac{F'(z)}{F(z)} = 2\pi i (N_{\rm z}- N_{\rm p})\,,
\end{equation}
where $N_{\rm z}$ and $N_{\rm p}$ are the number of zeros and poles enclosed in the contour. In our case, $\mD(\omega)$ is analytic and pole-free in the upper half-plane. Thus, applying the argument principle,
\begin{equation}
\frac{1}{2\pi i}\ointctrclockwise_{\mCw}\,d\omega \dfrac{\mD'(\omega)}{\mD(\omega)} = \frac{1}{2\pi}\ointctrclockwise_{\mCd}d\Big(\text{arg}\big[\mD(\omega)\big]\Big) \geq 1\,,
\label{eq:AP}
\end{equation}
which guarantees $N_{\rm z} \geq 1$ inside $\mCw$. Thus, under Conditions 1a, 1b, 2a, and 2b, there is at least one root of $\mD(\omega)$ in the upper half-plane. This completes the proof $\blacksquare$

\begin{figure*}
    \includegraphics[width=0.9\linewidth]{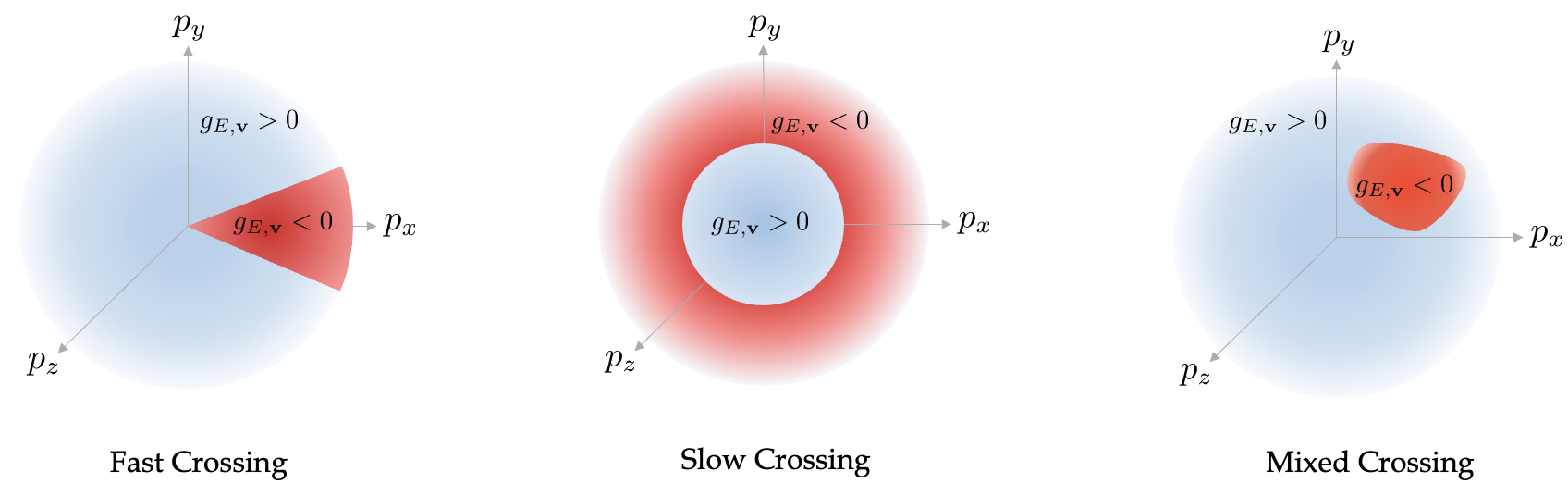}
    \caption{Illustrations of different types of crossing surfaces. The sign of the distribution $g_\bG$ is shown as blue ($>0$) or red ($<0$), as a function of $\bp=|\Delta m^2/(2\we)|\bv $ for one half of the allowed range of $\we$, say $\we\in(0,\infty)$. Crossings in only $\bv$ (left panel) are ``fast'' and unstable. Crossings in $E=|\bp|$ (or $\we$) but not $\bv$, are ``slow'' (middle panel); these are unstable under further conditions, i.e., on sign of $\partial g/\partial \we$ and on $\mI_\text{PV}$. The general case of ``mixed'' crossings is shown on the right panel, which is unstable only under further conditions.\label{fig:crossings}}
\end{figure*}

\subsection{Necessary Condition and Proof} 
\label{sec:prop_neces_cond}

Are all of these conditions necessary? Note that it is possible for $\mCd$ to encircle the origin even if Condition 2b is violated, i.e., with multiple crossings to the left of the origin but nevertheless encircling the origin. However, instability is then no longer \emph{guaranteed} by our approach. Thus, while the proposed conditions are sufficient, they are not strictly necessary. On the other hand, Conditions 1a, 1b, and 2a are indeed strictly necessary. This is easy to prove: for counter-clockwise winding around the origin (necessary for instability), the left-most crossing of $\mCd$ must be down-going and left of the origin (which embody the conditions 1a, 1b, 2a) $\blacksquare$

\subsection{Comments} 
\label{sec:comments}

A few remarks are in order. The conditions need to be adapted to the mass ordering and the causal character of the eigenvector. For definiteness, we have assumed inverted mass ordering ($\Delta m^2 > 0$ ) and time-like eigenvectors ($1-|{\bf a}|^2 >0$) in our discussion. For normal mass ordering, i.e. $\Delta m^2 < 0$, the measure of the integral in Eq.\,(\ref{eq:Pidef}), which has a factor of $(\Delta m^2)^3$, acquires a net minus sign. However, Eq.\,(\ref{eq:Pidef}) is invariant under the joint transformations $\Delta m^2 \to -\Delta m^2$ and $g_\bG\to-g_\bG$. Thus, the result for $\Delta m^2 < 0$ is the same as for the $\Delta m^2 > 0$ case but with an effective distribution $-g_\bG$. 
Likewise, for light-like eigenvectors with $|{\bf a}|^2=1$, the proof goes through without change, as one can simply drop the ``1" in Eq.\,(\ref{eq:disprel}); here $\mI_\text{PV}(\omega_0)<0$ with $\mI_\text{PV}(\omega_{i\neq0})>0$ is sufficient for winding around the origin.  For space-like eigenvectors with $1-|{\bf a}|^2 <0$ in Eq.\,(\ref{eq:scDR}), one can similarly absorb an overall minus sign in $g_\bG$ and find the corresponding conditions. Sec.\,\ref{sec:cnxns} and Sec.\,\ref{sec:summ} has more details.

It is also worth noting that, in general, Conditions 2a and 2b depend on the eigenvector in a nontrivial manner, through the velocity-dependent factor of $|1-{\bf a}\cdot{\bf v}|^2$, and not on $g_\bG$ alone. We will show later that for a large collective potential, $\mu \gg \we$, Conditions 2a and 2b will approximately soften to a weaker Condition 2 [see Eq.\,(\ref{eq:cond2})] that simply requires the smallest principal value to be negative. Considering that the above velocity-dependent factor has a fixed sign, Conditions 1b and  2 (as opposed to 2a, 2b) that simply restrict a sign and not magnitude, are affected only by the causal character of the eigenvector (not the full eigenvector).  In this limit, detailed dependence on the eigenvector softens to only a dependence on its causal character, and the necessary and sufficient conditions apply to the actual $g_\bG$ (without the above velocity-dependent factor).

\section{Relaxing the Conditions}
\label{sec:relaxing}

Conditions 1a and 1b impose local constraints on $g_\bG$, requiring a positive zero-crossing. Intuitively, such local information alone should not generally guarantee a global property, such as instability, that depends on an integral over all of $g_\bG$. Conditions 2a and 2b constrain the global structure of $g_\bG$. Condition 2a is needed to ensure that the contour $\mCd$ encircles the origin. Condition 2b is a further global constraint that prevents $\mCd$ from evading the origin due to the influence of other crossings. We now analyze whether some of these conditions can be softened or fully relaxed in special limits, e.g., fast or slow oscillations. 

\subsection{Positivity of Slope}
\label{sec:positivity}

Condition\,1b is the requirement that $(\nabla_\bG g_\bG)_0 \cdot (\nabla_\bG h_\bG)_0 >0$. Expanding this out, in terms of the gradient of $g_\bG$, one gets 
\begin{equation}
\Big(\bk - (\bk\cdot\bv_0)\bv_0\Big)\cdot\Big(\nabla_\bv g_\bG\Big)_0 > -\we_{_0}^2\left(\frac{\partial g_\bG}{\partial\we}\right)_0\,.
\label{eq:positivity}
\end{equation}
We now analyze this ``positivity'' condition in special cases of crossings, as illustrated in Fig.\,\ref{fig:crossings}.

\subsubsection{Fast Case}
\label{sec:fastcross}

Consider the case that $g_\bG$ is only a function of $\bv$, and not $\we$ (or $E$), and one works in the limit $\we \to 0$. For such situations, all crossings are purely ``fast crossings,'' and the right-hand side of Eq.\,(\ref{eq:positivity}) vanishes. Then the above condition simply requires the transverse part of $\bk$ to have positive projection on $(\nabla_\bv g_\bv)_0$. On $\bv\in\mathbb{S}^2$, the angular gradient $\grad_\bv g_\bv$ is orthogonal to $\bv$, and therefore \emph{any} $\bk$ chosen with a positive component along $(\grad_\bv g_\bv)_0$ satisfies the above inequality. Thus, for a fast crossing, the positivity condition on the slope is always satisfiable. This explains why for fast instability, e.g., in Morinaga's proof~\cite{Morinaga:2021vmc}, the slope at the zero-crossing does not need to bear a specific sign.

\subsubsection{Slow Case}
\label{sec:slowcross}

If we consider a distribution $g_\bG$ that is only a function of $\we$ (or $E$) but not $\bv$, one has $(\nabla_\bv g)_0=0$. For such purely ``slow crossings,'' the left-hand side of Eq.\,(\ref{eq:positivity}) vanishes and the positivity condition has nontrivial content: it can be satisfied for arbitrary $\bk$, but \emph{only if}
\begin{equation}
\left(\frac{\partial g_\we}{\partial\we}\right)_0>0\,,
\label{eq:slowpos}
\end{equation} 
i.e., $g_\we$ has a positive slope with respect to $\we$. Note that if $\Delta m^2 <0$ were considered instead, this condition would require a negative-slope crossing for instability. This explains the pattern observed for multiple spectral splits in Ref.~\cite{Dasgupta:2009mg}.

\subsubsection{Mixed Case}
\label{sec:mixcross}

For a ``mixed'' crossing, we consider that both $(\nabla_\bv g_\bG)_0$ and $({\partial g_\bG}/{\partial\we})_0$ are nonzero. If $({\partial g_\bG}/{\partial\we})_0<0$, Eq.\,(\ref{eq:positivity}) can only be satisfied if $\bk$ has a minimum projection
\begin{equation}
k_{\rm min} \equiv -\frac{\we_{_0}^2}{|(\nabla_\bv g_\bG)_0|}\Bigg(\frac{\partial g_\bG}{\partial\we}\Bigg)_0\,,
\label{eq:kminmixed}
\end{equation}
on $(\nabla_\bv g_\bG)_0$. In other words, if the distribution changes much more sharply along $\we$ (with negative slope), than it does along $\bv$, at all the crossing points, then instability is possible only for large wave-vectors with $|\bk|>k_\text{min}$. Of course, if $({\partial g_\bG}/{\partial\we})_0>0$, one has $k_\text{min}<0$ and the constraint is trivial.

\subsection{Negativity of Principal Value}
\label{sec:negativity}

Conditions\,2a and 2b, viz., $\mI_\text{PV}(\omega_0)<-1$ at some $\omega_0$ and $\mI_\text{PV}(\omega_{i\neq0})>-1$ at all other $\omega_i$, demand sufficient negativity of the principal value at one of the crossings. This can be significantly relaxed in some limiting cases. Below, we investigate when and how.

\subsubsection{Fast Case}
\label{sec:fastloose}

In the fast limit (i.e., neglecting $\we$ in the denominator, and then integrating out $\we$), the dispersion relation takes the simplified form
\begin{equation}
\mD_{\rm fast}(\omega,\bk)=1+\int d\bv \frac{G_\bv}{\omega-\bk\cdot\bv}=0\,,
\label{eq:fastdisp}
\end{equation}
where we have defined $G_\bv$ as the effective distribution on $\bv$ alone, as $\int d\bG\,g_\bG \equiv \int d\bv\,G_\bv$.

In this case, the principal value integrals in Conditions\,2a and 2b scale with the magnitude of $\bk$. Under $\bk \to \lambda\, \bk$, in Eq.\,(\ref{eq:fastdisp}) each singularity at $\omega_i$ shifts to $\lambda\omega_i$, and one has
\begin{equation}
 \mI^{\rm fast}_{\rm PV} \to \frac{1}{\lambda}\,\mI^{\rm fast}_{\rm PV}\,.
\label{eq:lamscal}
\end{equation}

As a result, it is sufficient to demand that $\mI_{\rm PV}(\omega_0)<0$ for some $\bk$, and one can make it more negative by simply letting $\lambda\to0^{+}$. We term this as the ``$\lambda$-scaling'' property of $\mI_{\rm PV}$.

To be precise consider that, for some $\bk$,  the smallest principal value among all $\mD$-crossings is $\mI_{\rm PV}(\omega_0)\equiv\kappa_0$ and the next largest is $\mI_{\rm PV}(\omega_1)\equiv\kappa_1 > \kappa_0$. If $\kappa_0<-1$ and $\kappa_1>-1$, then Conditions 2a and 2b are already satisfied. Suppose that, either $\kappa_0$ is not $<-1$, or that $\kappa_0<-1$ but $\kappa_1<-1$ as well. For $\bk\to \lambda\bk$, one has $\kappa_{i}\to \kappa_{i}/\lambda$ and Conditions 2a and 2b are satisfied for 

\begin{equation}
{\rm max}\left(0,-\kappa_1\right)\equiv \lambda^{\rm fast}_{\rm min} < \lambda < \lambda^{\rm fast}_{\rm max} \equiv -{\rm min}\left(0,\kappa_0\right)\,.
\label{eq:lamlim}
\end{equation} 
Given $\kappa_1>\kappa_0$, the above has solutions for $\lambda$ iff $\kappa_0<0$, i.e., the smallest $\mI_\text{PV}$ is negative.

In the fast limit, given a crossing at some $\bv_0$, an instability is therefore guaranteed for {some} wave-vector under a weaker condition that replaces Conditions 2a and 2b:
\begin{equation}
{\textit{\bf Condition}}\,\textbf{\textit{2:}}\quad \mI_{\rm PV}(\omega_0) = \pv \int d\bv\,\frac{G_\bv}{\bk\cdot(\bv_0 - \bv)} < 0\,.
\label{eq:cond2}
\end{equation}
Choosing $\bk_\perp$ along $\nabla_\bv G_\bv$ at the crossing-point gives the best chance of finding an instability: this choice satisfies positivity by construction and generates the most negative local contribution to $\mI_{\rm PV}$ at the crossing. One may worry that the far-contribution is not necessarily sub-dominant and may upend the sign. But if that is the case, one can choose $\bk'=-\lambda\bk$ with $\lambda>0$, so that $\mI_{\rm PV}(\omega_0)$ flips sign. This comes at the cost of $\bk'$ being anti-aligned to $(\nabla_\bv G_\bv)_0$, which is however not a problem, because the crossing curves on $\mathbb{S}^2$ are closed and we can consider a point $\bv_{0'}$ on the ``other side'' of the crossing-curve where $\bk'$ has a positive projection on the local gradient $(\nabla_\bv G_\bv)_{0'}$ and where $-\lambda\bk\cdot\bv_{0'} =\omega_0$. A simpler way to see this is that $\Delta m^2\to-\Delta m^2$ together with $g_\bG\to -g_\bG$, keeps Eq.\,(\ref{eq:disprel}) invariant; this means that the sign of $G_\bv$ is immaterial in the fast $\Delta m^2\to 0$ limit, and the principal value can be made to take both signs as long as it is not identically zero.

All in all, given a crossing and for fast collective oscillations described by Eq.\,(\ref{eq:EOM}) in the limit $\Delta m^2 \to 0$, Condition~1b and Condition 2 (or equivalently 2a,\,2b) are guaranteed to be satisfied for some $\bk$. One therefore has a remarkably simple condition for fast instability: ``the distribution $G_\bv$ must take both signs,'' precisely as was proved by Morinaga~\cite{Morinaga:2021vmc}.

 \subsubsection{Slow Case}
 \label{sec:slowloose}

In the slow limit, one considers that $g_\bG$ has no dependence on $\bv$, and the dispersion simplifies to
\begin{equation}
\mD_{\rm slow}(\omega,\bk)=1+\int d\bG \frac{g_\we}{\omega-\bk\cdot\bv-\we}=0\,.
\end{equation}
The denominator now has an absolute scale inherent in $\we$, and the principal values $\mI_\text{PV}(\omega_i)$ do not scale with $1/k$. Conditions\,2a and 2b cannot be simplified in this case. That is, e.g., if one has $\mI_\text{PV}(\omega_0)<0$ but not $<-1$ at some $\bk$,  there is no way to systematically choose a different $\bk$ to shift the crossing to the left. For small $\we$ there is an approximate version of $\lambda$-scaling, that we will discuss shortly.

For homogenous slow modes with $\bk=0$, there is no $\bk$ left to scale with. Here $\omega_0=\we_{_0}$, and the Condition 2a is
\begin{equation}
\mI_\text{PV}(\omega_0) \equiv \int d\we \frac{G_\we}{\omega_0-\we}<-1\,,
\end{equation}
with $\int d\bG\,g_\we \equiv \int d\we\,G_\we$. This inequality bears some resemblance to the second ``consistency condition'' for single-angle slow instabilities; see, e.g., Eq.(37) of Ref.\,\cite{Banerjee:2011fj}. In this context, we believe Condition\,2b on the principal values at $\omega_i = \we_{_i}$, corresponding to other crossings,
\begin{equation}
\mI_\text{PV}(\omega_i) \equiv \int d\we\,\frac{G_\we}{(\omega_i-\we)}>-1\quad\text{for}~i\neq0\,,
\end{equation}
has not been noted before. Without this, encirclement and instability are not guaranteed, especially for contrived distribution functions with large higher derivatives.

\subsubsection{Mixed Case}
\label{sec:mixloose}

Finally, for the mixed case where we allow $g_\bG$ to depend on both $\we$ and $\bv$ at the crossing, again, there is no strict scaling with $\bk$. However, for sufficiently small $\we$, as realized in physical environments such as core-collapse supernovae, where $\langle \we \rangle \sim {\cal O}(10^{-5})$ in units of $\mu=1$, an approximate form of the $\lambda$-scaling emerges. In this regime, Conditions 2a and 2b are expected to soften considerably, potentially resembling Condition 2 up to small corrections. We proceed to quantify this behavior.

\pagebreak

The principal value integral
\begin{align}
    \mI_{\rm PV}(\omega_i) &= \pv\int d\bG\;\dfrac{g_{\bG}}{\omega_i - \bk\cdot\bv-\we}\,
\end{align}
is parameterized by a physical constant, namely $\Delta m^2$, and we want to qualitatively understand the behavior of the integral in the limit $\Delta m^2\to0$. It is worth noting that this dependence enters explicitly via the $\we$ in the denominator, and the factors of $\Delta m^2$ cancel elsewhere (e.g., in $d\bG$).

The crossing frequencies $\omega_i$ themselves depend on the small parameter $\Delta m^2$. To first order, we take
\begin{equation}
    \omega_i = \omega^{\rm fast}_i + \omega^{\rm \Delta}_i\,,
\end{equation}
i.e., $\omega_i$ differs by an order-$\Delta m^2$ shift, denoted as $\omega^{\rm \Delta}_i$, from its value in the fast limit. Computing the principal value integral to first order in $\Delta m^2$, we find
\begin{equation}
    \mI_{\rm PV}(\omega_i) = \mI_{\rm PV}^{\rm fast} + \mI_{\rm PV}^\Delta\,,
\label{eq:IPVIDelta}
\end{equation}
where
\begin{align}
\mI_{\rm PV}^{\rm fast}&=  \pv\int d\bG \frac{g_\bG}{\omega^{\rm fast}_i - \bk\cdot\bv}\,,\\
\mI_{\rm PV}^\Delta&= \pv\int d\bG\, \frac{\big(\we-\omega_i^\Delta\big)\,g_\bG}{(\omega^{\rm fast}_i - \bk\cdot\bv)^2}\,,
\end{align}
with $\mI_{\rm PV}^\Delta$ being of order $\Delta m^2$. This result is obtained under the assumption that $g_\bG$ vanishes at $|\bv|=1$, as is expected for massive neutrinos with $|\bv|=|{\bf p}|/|E| <1$. See the Appendix for a detailed proof and discussion. 

Under $\bk \to \lambda\bk$ and $\omega_i^{\rm fast}\to\lambda\omega_i^{\rm fast}$, to linear order in $\Delta m^2$, we  thus have
\begin{equation}
    \mI_{\rm PV}(\omega_i) \to  \frac{\mI_{\rm PV}^{\rm fast}}{\lambda} + \frac{\mI_{\rm PV}^\Delta}{\lambda^2}\,.
\end{equation}
That is, one can effect an approximate multiplicative change to $\mI_{\rm PV}$ by $\lambda$-scaling, as long as  the $\mI_{\rm PV}^\Delta$ term remains small. Scaling $\lambda$ to larger than 1 imposes no additional restriction from a nonzero $\Delta m^2$, as it makes the $1/\lambda$ scaling more exact. But if one chooses $\lambda\to0^+$, e.g., so that a small negative $\mI_{\rm PV}$ at the leftmost crossing at $\omega_0$ can be made $<-1$, one can only do so to the extent that ${\mI_{\rm PV}^\Delta}/{\lambda^2}\lesssim{\mI_{\rm PV}^{\rm fast}}/{\lambda}$, i.e., the non-scaling corrections remain small. In other words, we have an approximate $\lambda$-scaling behavior, over a range of $\lambda$,
\begin{equation}
{\rm max}\bigg( \tfrac{{\mI_{\rm PV}^\Delta(\omega_0)}}{\mI^{\rm fast}_{\rm PV}(\omega_0)} , \lambda_{\rm min}^{\rm fast}\bigg) \lesssim \lambda \lesssim \lambda_{\rm max}^{\rm fast}\,,
\label{eq:welamlim}
\end{equation}
that is slightly reduced at the lower end, relative to the range in Eq.\,(\ref{eq:lamlim}), by the $\Delta m^2$-dependent corrections.

This approximate $\lambda$-scaling property of $\mI_\text{PV}(\omega_i)$ suggests that Conditions 2a and 2b can usually be satisfied as long as the weaker version, Condition 2, is satisfied, unless they are violated so strongly to require $\lambda$ outside the above range. Arguments similar to the fast case suggest that one can satisfy Condition 2, because the crossing-surfaces are closed and one can choose the crossing point and direction of $\bk$ appropriately.

Still, there is one possible roadblock. If  $-(\partial g_\bG/\partial \we)_0\gg |\nabla_\bv g_\bG|_0/\we_{_0}^2$ at all crossings, the minimal wave-vector $k_\text{min}$, see Eq.\,(\ref{eq:kminmixed}), becomes very large and may conflict with satisfying Condition\,2a by scaling down $\bk$, even if it is within the allowed range of approximate $\lambda$-scaling in Eq.\,(\ref{eq:welamlim}). In such a case, instability is absent.

Practically, as long as scaling is expected to be successful, a positive $\we$-slope crossing implies instability. If $(\partial g_\bG/\partial \we)_0<0$ but with $\we$-slopes not too negative and $\bv$-gradient not too small, instability is expected. This is because the $\lambda$-scaling property softens Conditions 2a and 2b to Condition 2, which is expected to be satisfiable with acceptable $\bk$ in the limit of small $\we$. A crossing is, in this sense, an \emph{almost sufficient} criterion for instability.

\section{Numerical Examples}
\label{sec:examples}

We present some examples to illustrate the conditions laid down in the proposition and show that encirclement is guaranteed only when all of them are satisfied.
\subsection{Fast Instability}
We begin by considering the fast oscillation limit, wherein the vacuum term $\Delta m^2$ vanishes and spectral crossings arise purely as a function of velocity. In this regime, the dispersion relation simplifies to the form given in Eq.\,(\ref{eq:fastdisp}), with the distribution function $g_{\bG}$ identified as $G_{\bv}$. As representative examples, we consider azimuthally symmetric distributions that are composed of one or more Gaussians in a variable $v$. The functional form of the distribution function is given by $G_{v} = \sum_{i} a_i \mathcal{N}_v(\mu_i,\sigma_i)$, where the $\mathcal{N}_v$ is a standard Gaussian in $v$, with mean $\mu_i$ and variance $\sigma_i$. 

In each figure panel, we display the distribution function $g_{\bG}$, the contour $\mCw$ in the complex $\omega$-plane, and the corresponding mapped contour $\mCd$ in the $\mathcal{D}$-plane. The mapping between $\mCw$ and $\mCd$ is indicated via a color gradient across both contours. The semicircular arc in the upper half of the $\omega$-plane is mapped to the single point $(1,0)$ on the real axis, which is highlighted in purple. As $\omega$ increases from $-\infty$ along the real axis, the contour $\mCd$ departs from $(1,0)$. The segment corresponding to $\omega \in (-k, +k)$, where $\mI_{\delta}\neq0$, is represented by a color gradient ranging from blue to red, indicating the upward movement of the contour and its approach toward $(1,0)$ from below as the imaginary part of the dispersion integral $\mI_{\delta}$ changes sign across a crossing.

\begin{figure}[]
    \centering
    \includegraphics[width=0.85\linewidth]{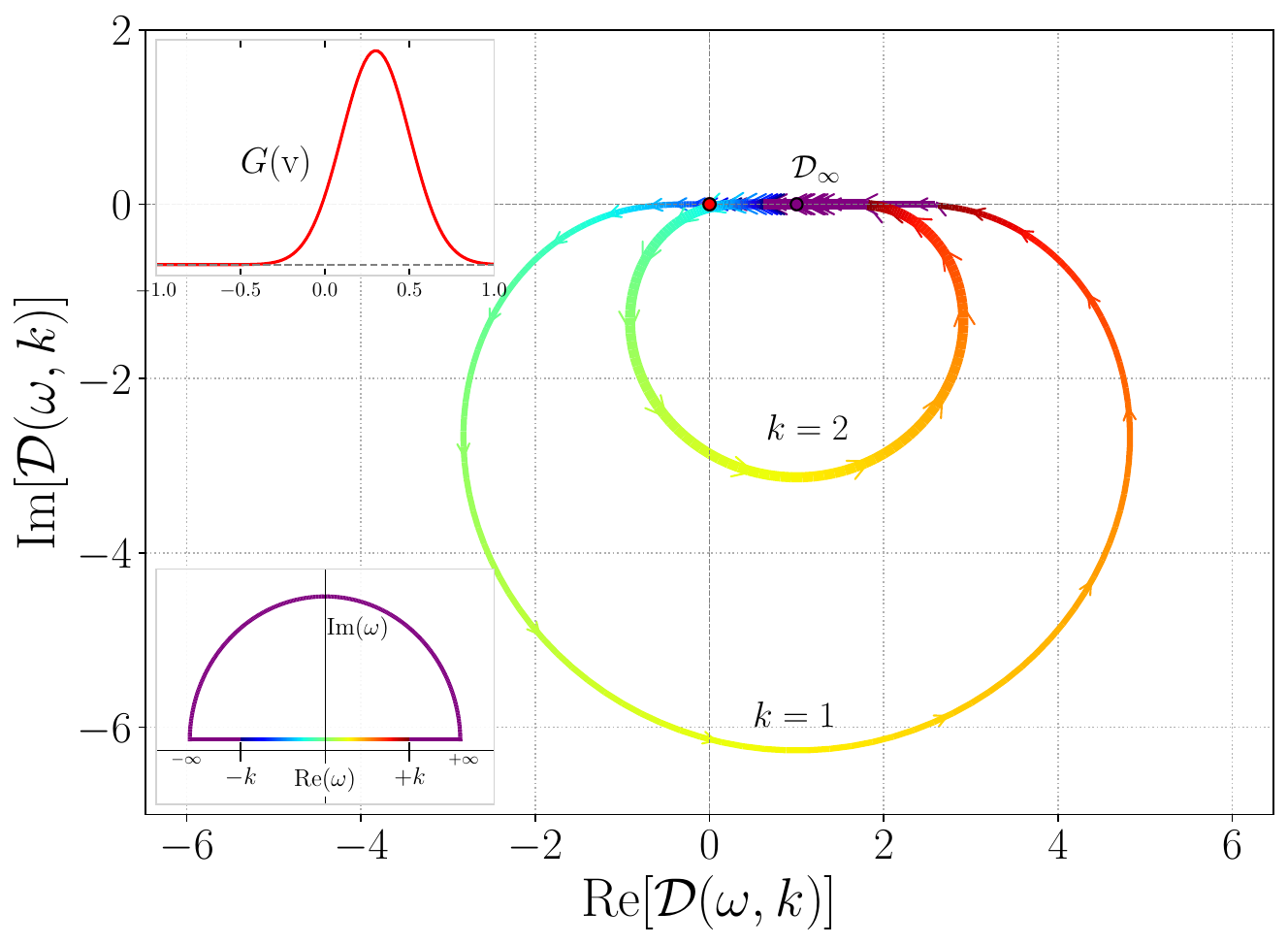}\vspace{0.3cm}
    \caption{Distribution function with no crossing results in no encirclement of origin for any value of $k$. Note how, for the fast case, increasing $k$ shrinks the contour without distortion of its shape.}
    \label{fig:gv_1Gaussian}
\end{figure}

\begin{figure}[]
    \centering
    \includegraphics[width=0.85\linewidth]{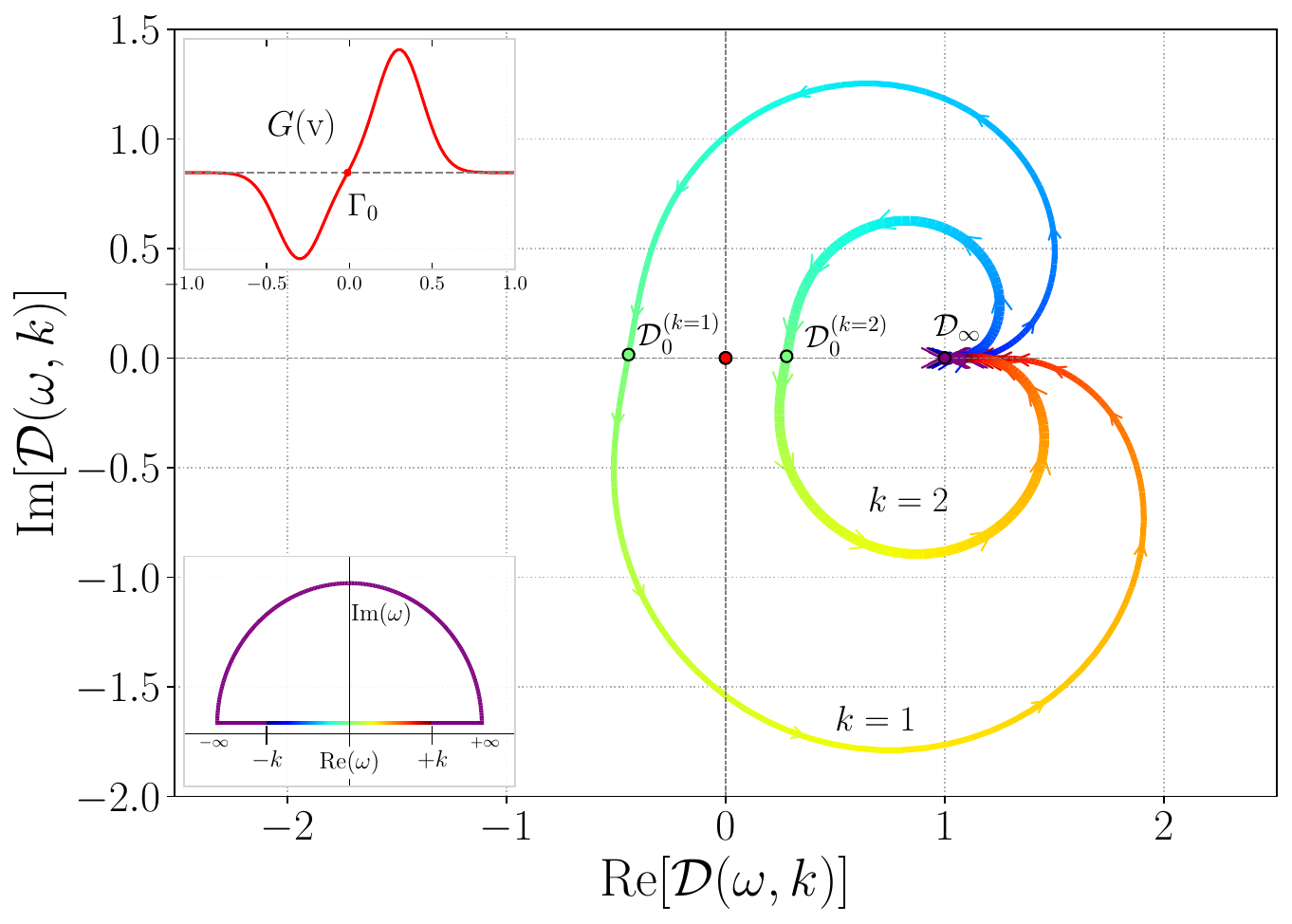}\vspace{0.3cm}
    \caption{Distribution function with a single crossing at $v=0$, that leads to two $\mCd$ crossings, at $\mD_0<0$ and at crossing at $\mD_\infty=1$, giving origin-encirclement. Instability exists in the range $\omega\leq|k|$ which is shown in the inset.  For large values of $k$, $\mCd$ shrinks and moves closer to $\mD_\infty$, eventually resulting in no encirclement.} 
    \label{fig:gv_2Gaussian}
\end{figure}

In Fig.\,\ref{fig:gv_1Gaussian}, the result is shown for a single Gaussian centered around $\mu=0.3$ and with a width $\sigma =0.2$. This violates Condition 1a, as there is no crossing, and clearly there is no encirclement of origin, and there is no instability for any $k$. By way of contrast, in Fig.\,\ref{fig:gv_2Gaussian} we show a case with a single crossing. The parameters are $(a_1, \mu_1, \sigma_1)=(-0.14,-0.3,0.14)$ for the Gaussian on the left, and $(a_2, \mu_2, \sigma_2)=(0.2,0.3,0.14)$ for the Gaussian on the right. In this case, $\mCd$ encircles origin for small $k$ and gives an instability, but as $k$ is increased $\mCd$ shrinks and avoids encirclement.


\begin{figure}[t]
    \centering
    \includegraphics[width=0.87\linewidth]{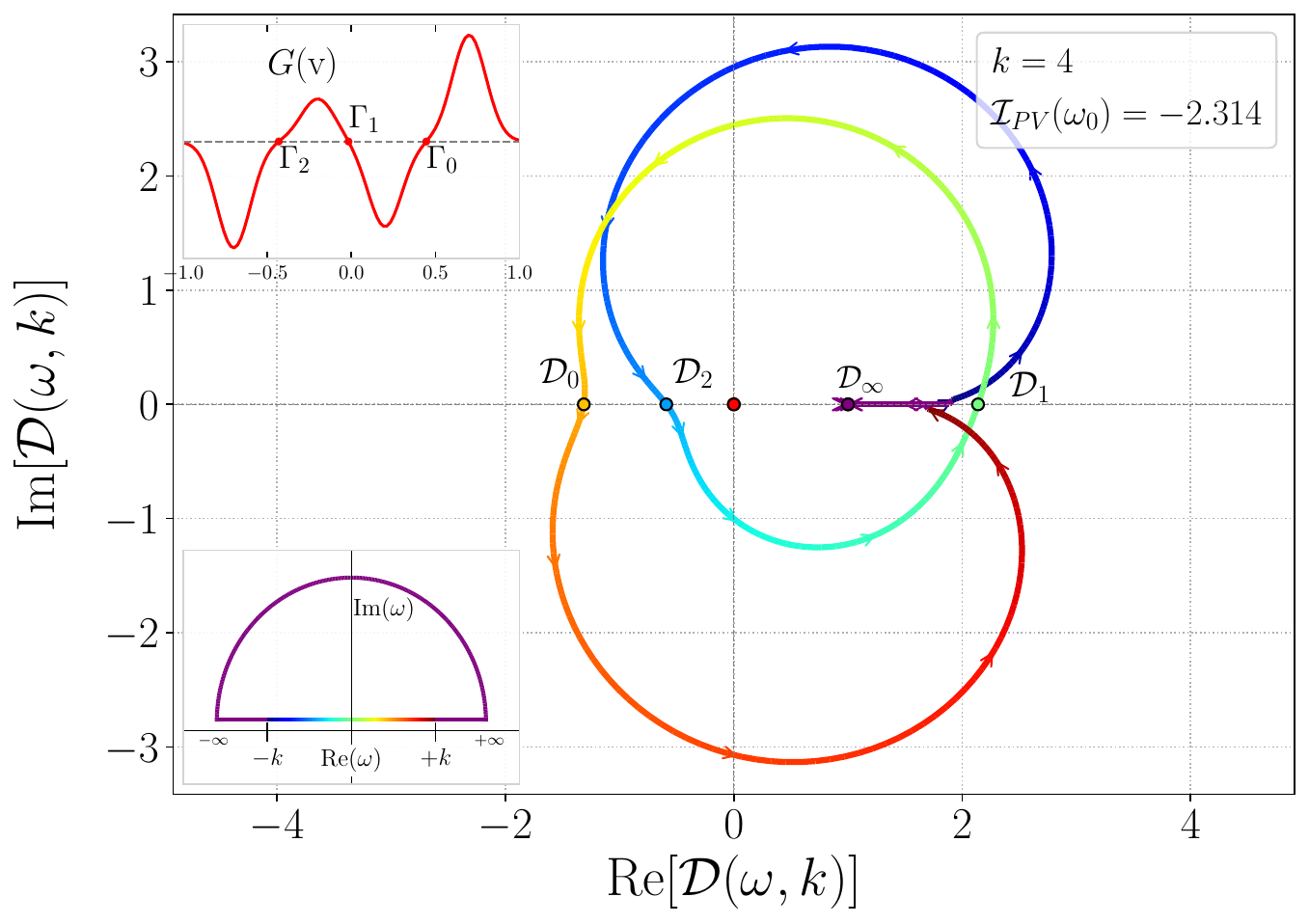}\vspace{0.6cm}
    \includegraphics[width=0.87\linewidth]{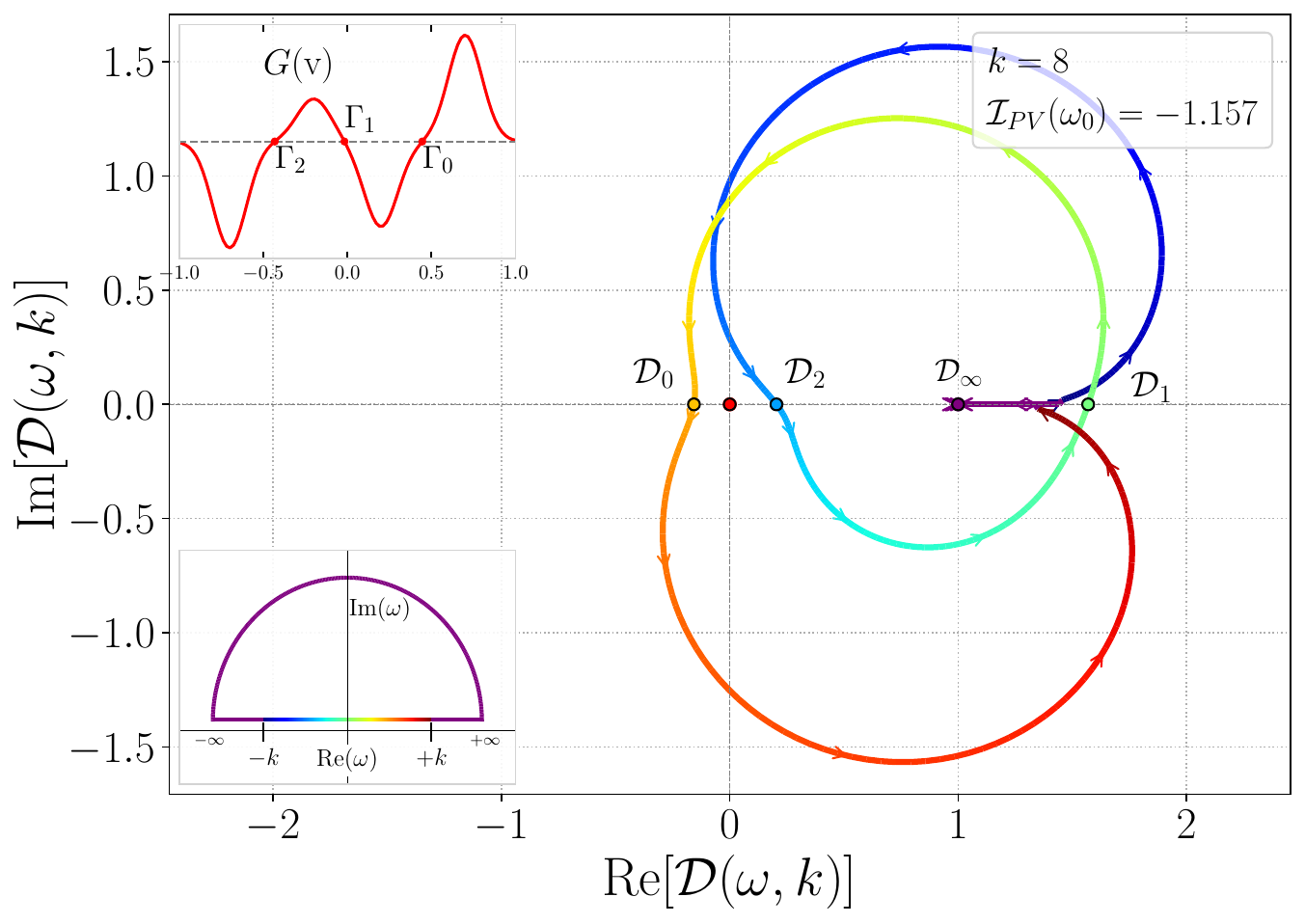}\vspace{0.6cm}
    \includegraphics[width=0.87\linewidth]{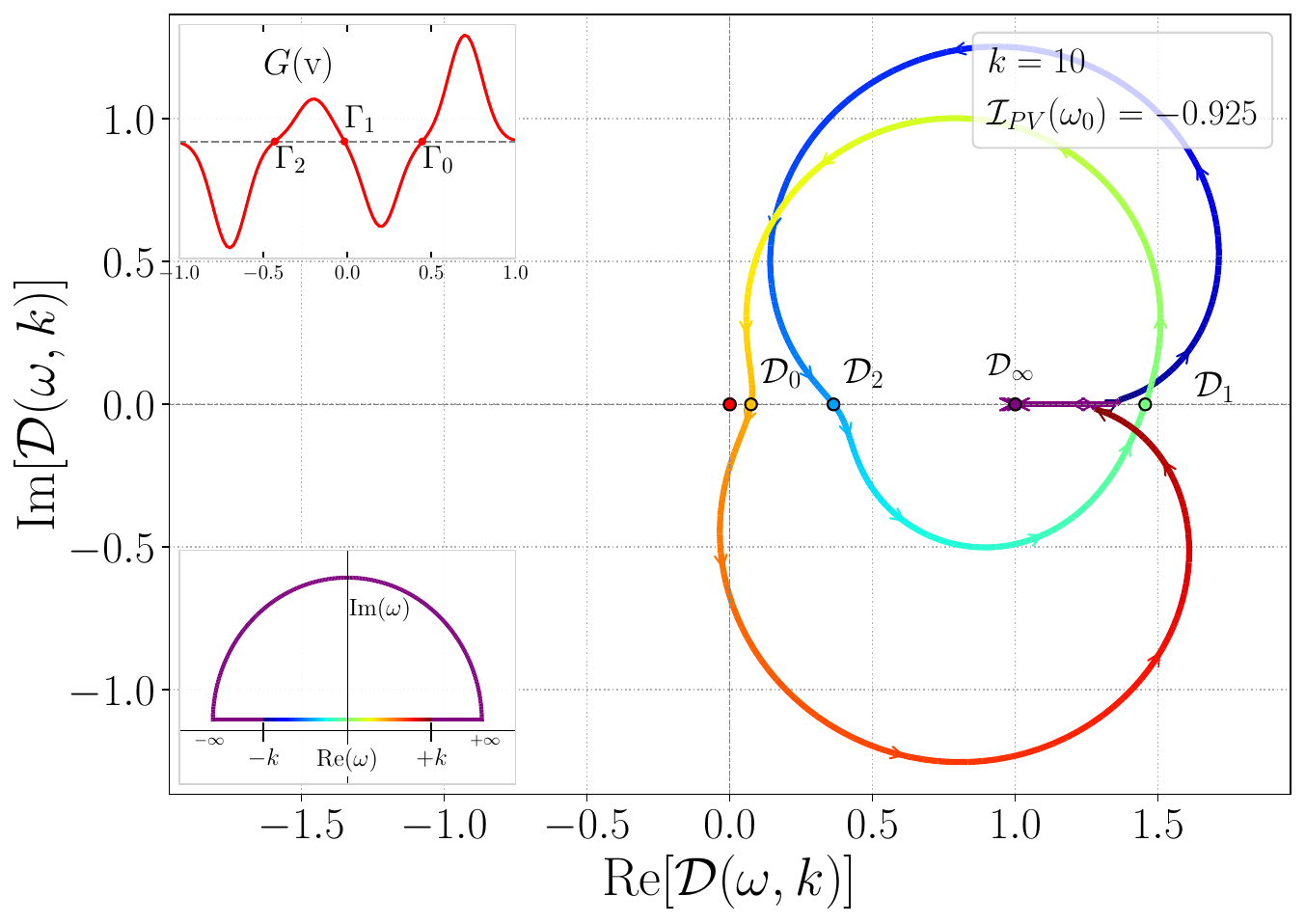}
    \caption{Triply-crossed distribution $G_v$ leads to 4 crossings of $\mCd$. The panels show the effect of scaling $k$, keeping $G_v$ the same. As one increases $k$, from top to bottom, the contours shift rightward as expected from $\lambda$-scaling. In this case, there is origin encirclement in the top and middle panels, but not in the bottom panel.}
    \label{fig:gv_4Gaussian}
\end{figure}

\begin{figure}
    \centering \vspace{-0.5mm}
    \includegraphics[width=0.85\linewidth]{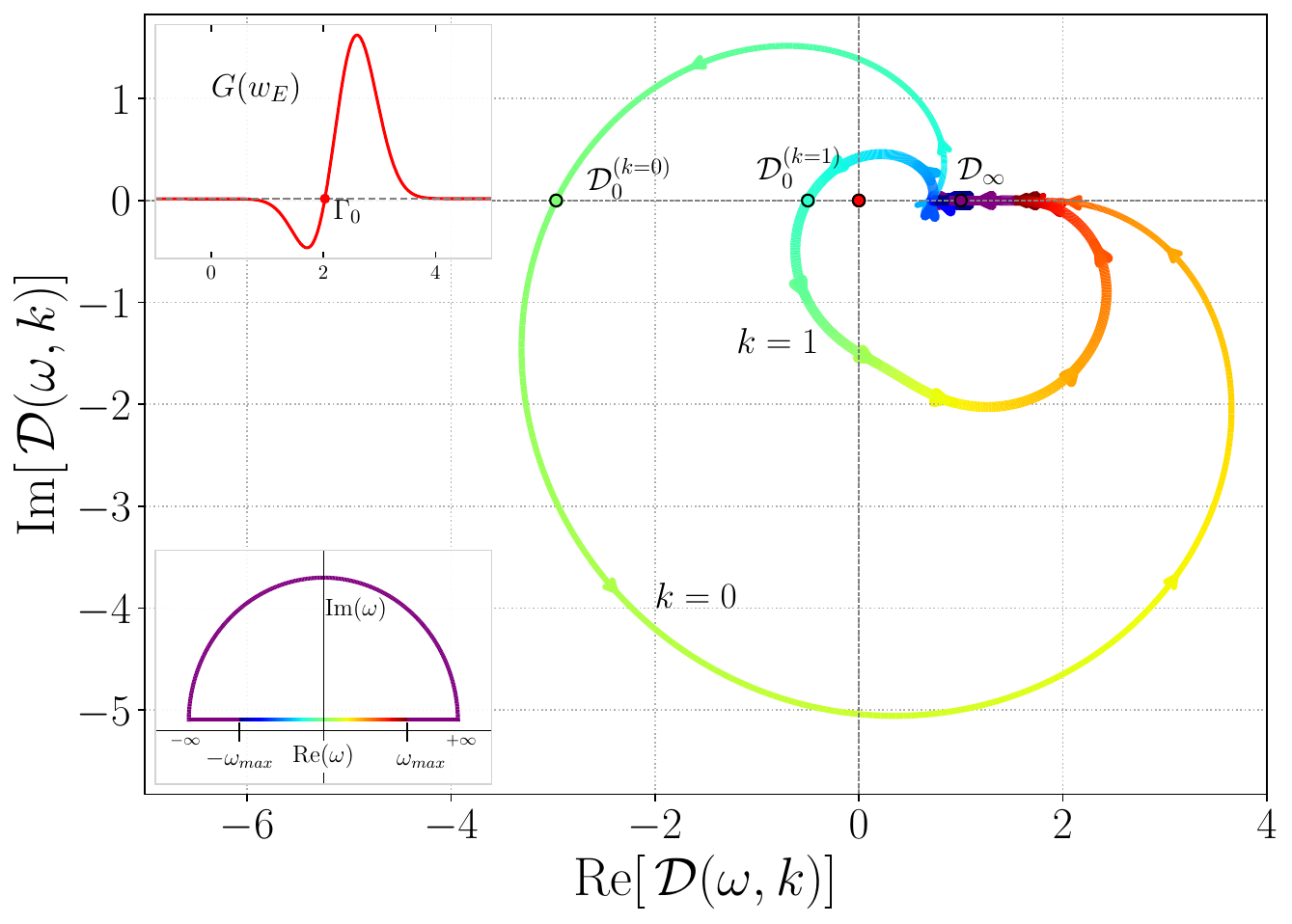}\vspace{0.5cm}
    \includegraphics[width=0.85\linewidth]{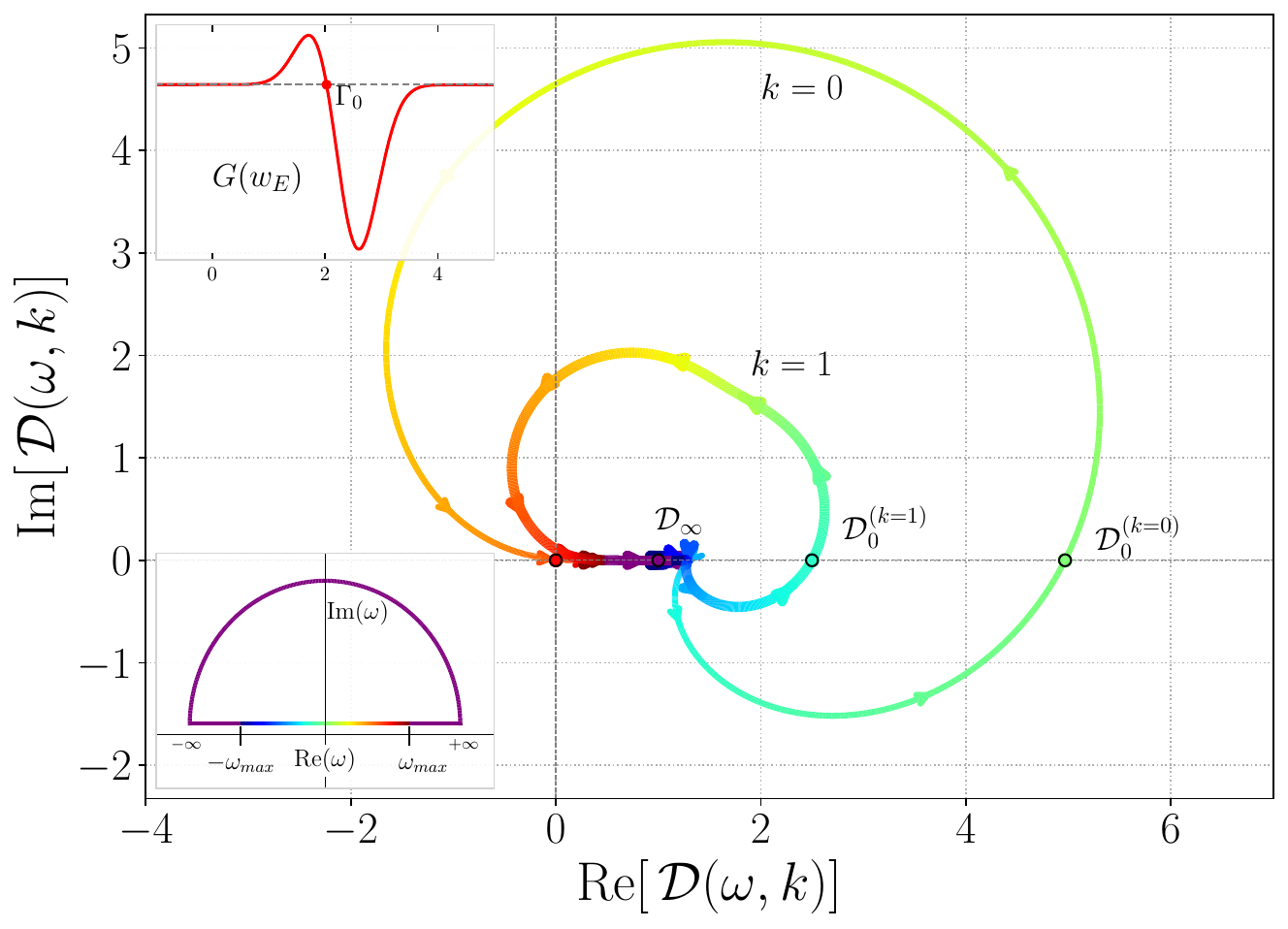}\vspace{0.5cm}
    \caption{Single crossing in $\we$. The panels show the effect of changing the distribution function $G_\we$ and $k$. The top panel shows a positive crossing, which can lead to origin-encirclement. The bottom one, with only a negative crossing, does not, illustrating Condition\,1b. Note also that scaling of $k$ distorts the $\mCd$ contour, in addition to scaling it. Thin (thick) curves correspond to $k=0$ ($k=1$). In the inset, $\omega_{\text{max}}$ refers to some $\omega$ beyond which there exists no instability. }
    \label{fig:gE_2gaussian}
\end{figure}

The significance of Conditions 2a and 2b is clearer in the case of multiple crossings, as shown in Fig.\,\ref{fig:gv_4Gaussian}. Here, we choose a set of four Gaussians with alternately opposite signs, with the following parameters for the Gaussians appearing from left to right, $(a_1, \mu_1, \sigma_1)=(-1,-0.7,0.1)$, $(a_2, \mu_2, \sigma_2)=(0.4,-0.2,0.1)$, $(a_3, \mu_3, \sigma_3)=(-0.8,0.2,0.1)$, and $(a_4, \mu_4, \sigma_4)=(1,0.7,0.1)$. Due to multiple crossings, the contour $\mCd$ exhibits multiple windings around the origin, as shown in the top panel. The principal value at the left-most $\mD$-crossing is indicated in the panels, and increases as $k$ increases. Interestingly, the middle panel shows a realization of successful $\lambda$-scaling, when only  $\mI_{\rm PV}(\omega_0) <-1$\, at a real $\omega_0$\,, while for the other crossings $\omega_i$, we have $\mI_{\rm PV}(\omega_i)>-1$. In this case, violating 2b (top panel) does not lead to avoidance of encirclement; this would typically require more contrived distribution functions where the higher derivatives dominate the lower derivatives.

\subsection{Slow Instability}
In this section, we present examples corresponding to the purely slow oscillation limit. Here, we plot the contours corresponding to the distribution function $G_\we$, defined via $\int d\bG\,g_{\bG} = \int d\we G_{\we}$. The distribution function is a sum of two Gaussians with the parameters $(a_1, \mu_1, \sigma_1)=(\mp0.5,2,0.4)$ and $(a_2, \mu_2, \sigma_2)=(\pm1,2.5,0.4)$, to have a positive/negative crossing along $\we$. In the panels of Fig.\,\ref{fig:gE_2gaussian}, we present these two opposing scenarios with positive or negative crossings along $\we$, each with two different values of $k$; thin (thick) lines correspond to $k=0$ ($k=1$). As is expected from Eq.\,(\ref{eq:slowpos}), for positive crossing along $\we$ (top panel) we find encirclement, and for negative crossings (bottom) we do not, illustrating the significance of Condition 1b. Note how scaling of $k$ has an effect of approximately scaling $\mD_0$, but with some distortion of shape as we predicted. In the slow-limit $g_\bG\to G_\we$, there is a $\bk\to-\bk$ symmetry, as well as invariance under $g\to-g$ with $(\mD-1) \to -(\mD-1)$.

\subsection{Mixed Instability}

\begin{figure}[t]
    \centering \vspace{-0.2cm}
    \includegraphics[width=0.92\linewidth]{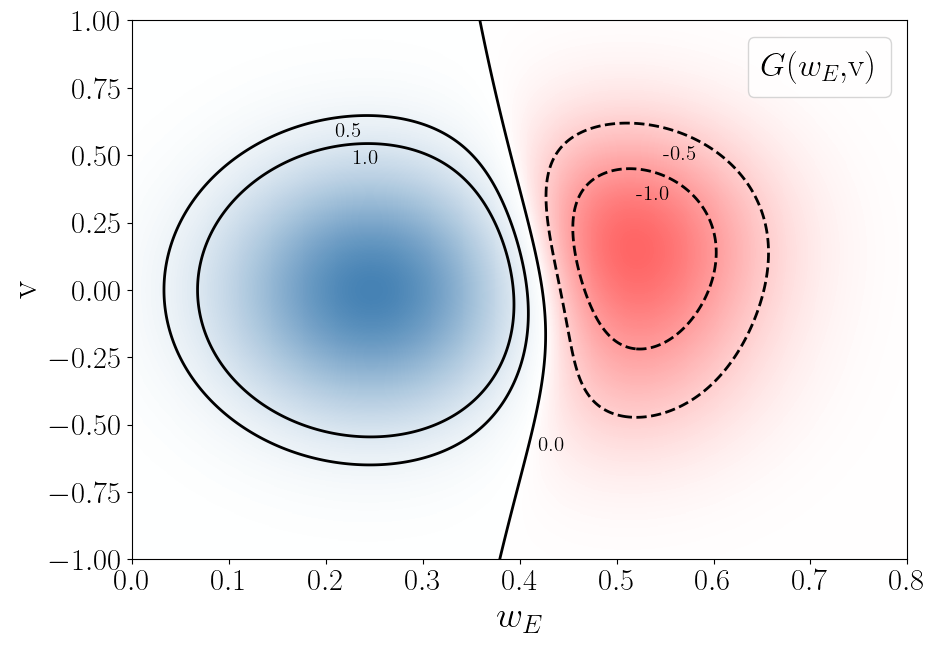}\vspace{0.4cm}
    \includegraphics[width=0.9\linewidth]{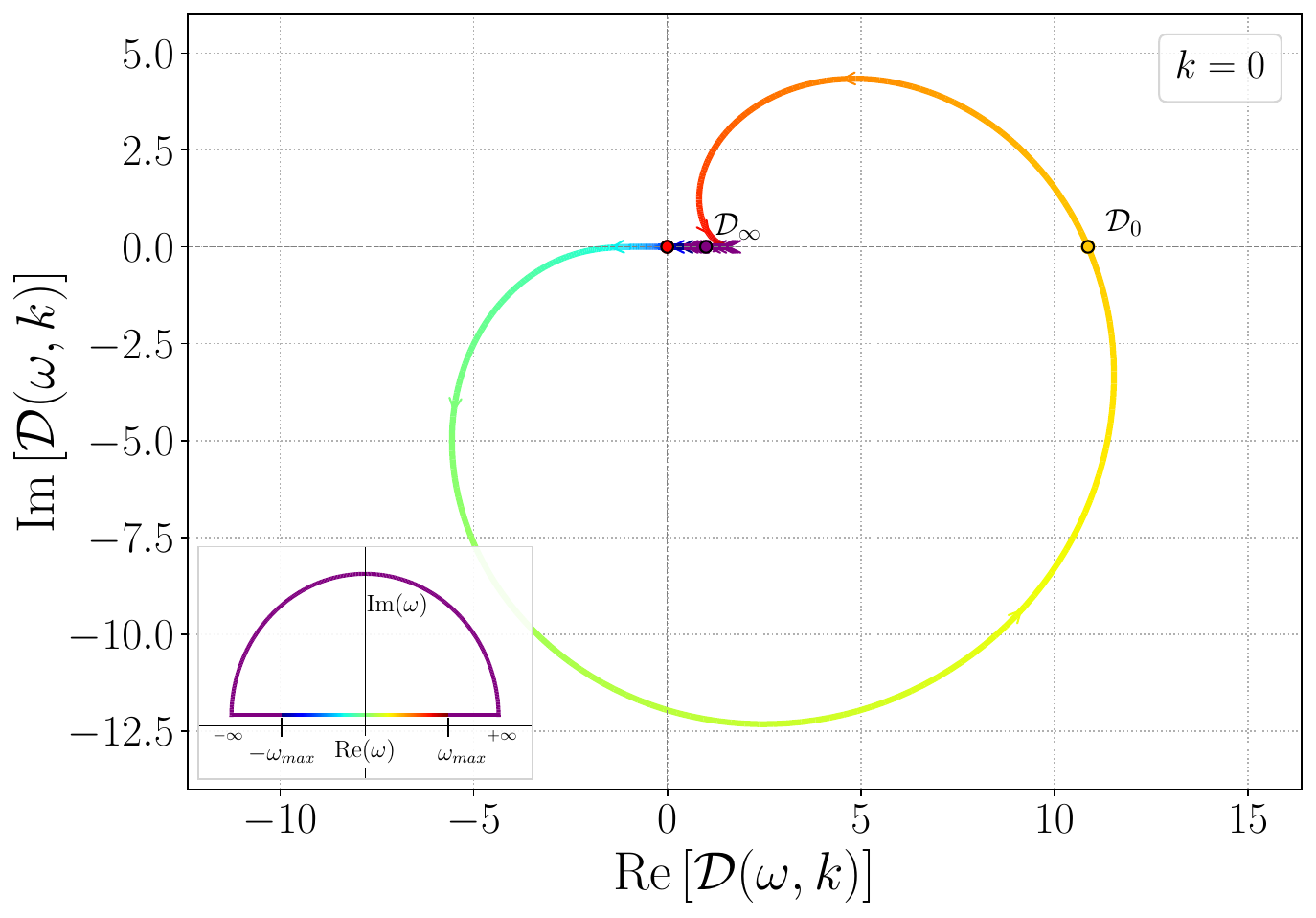}\vspace{0.4cm}
    \includegraphics[width=0.9\linewidth]{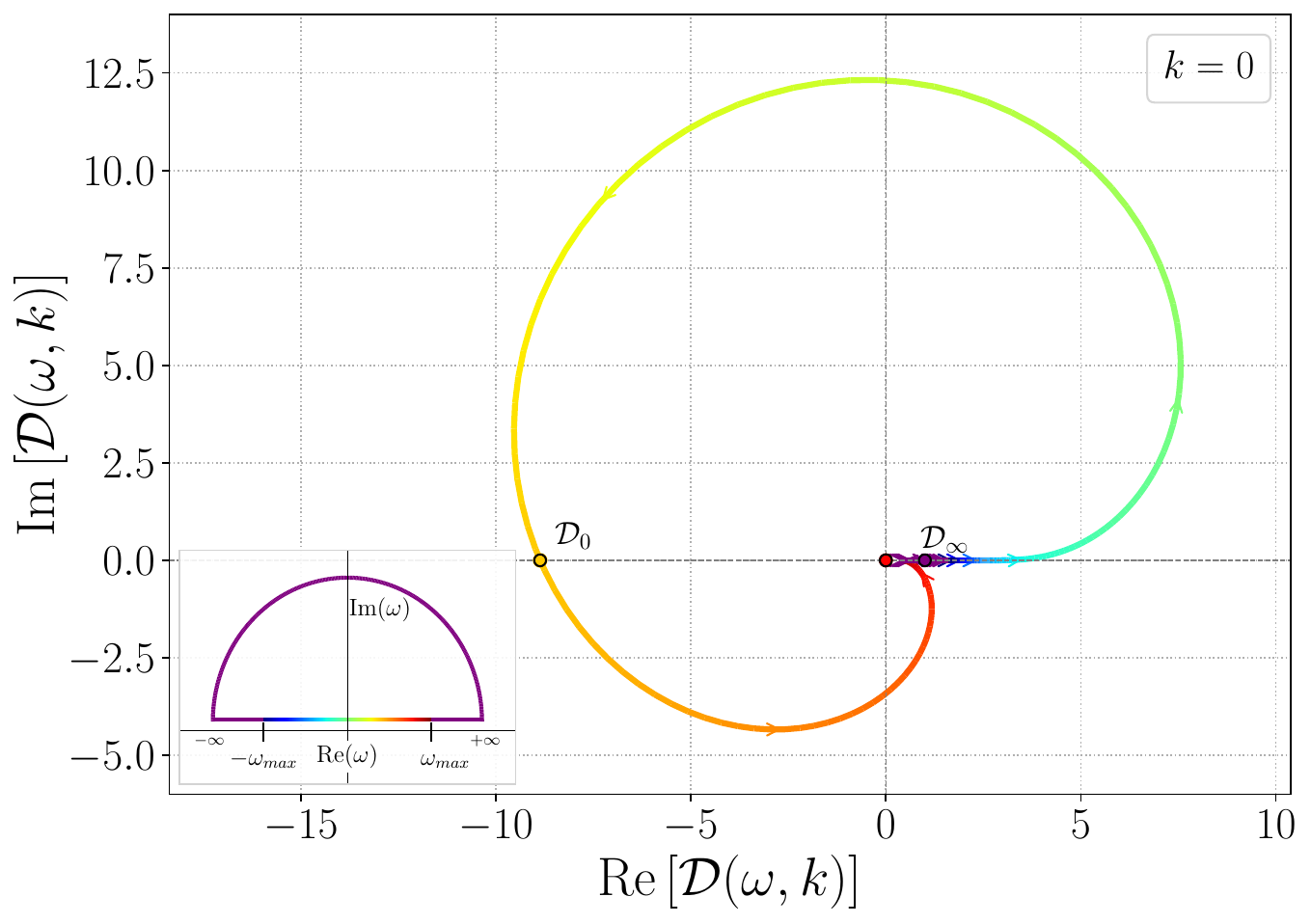}   
    \caption{Mixed crossing in $\we$ and $v$. The top panel shows the distribution function with only negative crossings along $\we$, thus there is no origin-encirclement for $k=0$ (middle panel). For the same $G$ but with the sign flipped, one finds origin-encirclement (bottom panel). \label{fig:gMixed}}
\end{figure}

Finally, we consider an example of the general case, where the distribution depends nontrivially on both $v$ and $\we$, e.g., $\int d\bG\,g_\bG = \int d\we\, dv\,G_{\we,v}$. In the top panel of Fig.\,\ref{fig:gMixed}, we show an azimuthally symmetric distribution function 
\begin{align}
 G(\we,v)=\mathcal{N}_{v}&(0,0.3)\,\Big[\mathcal{N}_{w_E}(0.25,0.1)\nonumber \\
 &- 0.3\,(1+v+3v^2)\mathcal{N}_{w_E}(0.5,0.1)\Big]\,,
\end{align}
that has crossings along both $\we$ and $v$. The distribution is a Gaussian around $v=0$ and falls off at $v\to\pm1$; it has a surface of zero-crossing points around $\we\approx0.4$, extending through all $v$ and closing off at the poles, with positive and negative Gaussian blobs at smaller and larger $\we$, respectively. The parameters have been chosen to ensure that $G(\we,v)$ does not have any symmetry along either of its coordinates.
As one can see, $\partial G/\partial \we <0$ at all points on the zero-crossing surface. We therefore expect, at least for $k=0$ there should not be any instability in this case. The middle panel of Fig.\,\ref{fig:gMixed} shows that indeed there is no winding around the origin for $k=0$. If, however, we flip the sign of $G$, there is winding (and thus instability) for $k=0$, as shown in the lower panel.

\section{Discussion and Summary}
\label{sec:discsum}
Before we conclude, we take a moment to situate our work in the context of previous work on plasma instabilities and the broad area of collective neutrino oscillations.
\subsection{Comparison with Penrose's Conditions}
\label{sec:penrose}
Our approach parallels and generalizes the well-known stability conditions for plasma waves, derived by Oliver Penrose~\cite{Penrose1960}. He considered a dispersion relation of the form
\begin{equation}
\mD_\text{\,P}(\omega,\bk) \equiv 1+\frac{4\pi n e^2}{k^2m}\int d\bv\frac{\bk\cdot({\partial f_\text{P}}/{\partial \bv})}{\omega-\bk\cdot\bv}=0\,,
\end{equation}
where we have adapted the original notation to avoid confusion with our case. For arbitrary \mbox{$\bk = k\,\hat{z}$}, and absorbing numerical factors into a redefinition of the distribution, the dispersion relation becomes
\begin{equation}
\mD_\text{\,P}(\omega,k) \equiv 1+\int dv\frac{f'(v)}{\omega-kv}=0\,.
\end{equation}
This bears clear similarities to the dispersion relation we have been working with. What we call $g$, appears as a function $f'= \partial f/\partial v$ in plasma context, shown above.

Penrose's conditions for instability (necessary and sufficient) are given by
\begin{itemize}
\item[{\bf\textit{P1:}}] $f(v)$ has a minimum at some $v=v_0$\,,~\text{and}
\item[{\bf\textit{P2:}}] $\pv\displaystyle\int dv \dfrac{f'(v)}{v_0-v} < 0$.
\end{itemize}
Our Conditions\,1a~(zero-crossing, given by $g=0$) and 1b~(positivity, given by $g'>0$) correspond to Condition P1 above ($f'=0$ and $f''>0$). Conditions\,2a and 2b, when softened to Condition\,2, i.e., $\pv\int d\bG g_\bG / (\bk\cdot(\bv_0-\bv))<0$, correspond to Condition\,P2 above. 

Our work has two key novelties. First, our analysis applies to ultra-relativistic momentum distributions over the three-dimensional $(\we,\bv)\in\mathbb{\Gamma}$ space, rather than to non-relativistic distributions over $\bv \in \mathbb{R}^3$ as in Penrose's case. We identified the appropriate generalization of the condition ``$f''>0$," namely, $(\nabla_\bG g_\bG)_0 \cdot (\nabla_\bG h_\bG)_0 > 0$ (Condition 1b). Second, Penrose's Condition P2 had to be strengthened: for $\we \neq 0$, an absolute frequency scale appears, and the factor $1/k$ can no longer be freely scaled to ensure that $\mCd$ winds around the origin. We replaced Condition P2 with Conditions 2a and 2b, which together guarantee winding and instability. 

Notably, the compact topology of $\mathbb{S}^2$ relaxes the requirements on crossing-slope and principal values in the fast limit, whereas the absolute scale introduced by $\Delta m^2$ requires a strengthening of Condition P2. These opposing effects make it nontrivial, a priori, to infer the correct conditions. In response, we returned to the more general Nyquist argument~\cite{Nyquist:1932}.

Finally, we note that in Penrose's case, Conditions P1 and P2 can be combined into a single sufficiency criterion: $f(v)$ must have a \emph{global} minimum at $v_0$. Condition P2, after integration by parts yields
\begin{equation} \pv\int dv\,\frac{f(v_0)-f(v)}{(v_0-v)^2}<0\,,
\end{equation}
which is satisfied if $v_0$ is a global minimum of $f$. This elegant result expresses a purely geometric condition on $f$, independent of any choice of wave-vector; cf. Eq.\,(\ref{eq:cond2}). In our general case, such a formulation does not appear to be possible.

\subsection{Connections to Related Work}
\label{sec:cnxns}
Apart from core-collapse supernovae, neutrino flavor transformations become crucial in other dense environments, e.g., neutron star mergers~\cite{Dasgupta:2008cu,Wu:2017qpc,Grohs:2022fyq,Froustey:2023skf}, and the early Universe~\cite{Froustey:2020mcq}. Inherently, the evolution of a dense neutrino gas is a model-system for many-body quantum dynamics with interesting entanglement features~\cite{Pehlivan:2011hp,Patwardhan:2022mxg,Cervia:2019res,Roggero:2022hpy}, and thus may also be of interest to the growing field of quantum information.

A more direct connection in the context of the present work is to plasma instabilities. Perhaps the first hint of this connection appeared in the development of linear stability analysis~\cite{Banerjee:2011fj, Izaguirre:2016gsx, Airen:2018nvp} that gave equations in the form that resembles the dispersion relations of plasma waves.  A more explicit connection was established in Ref.\,\cite{Capozzi:2017gqd} where, by analyzing the pole structure in the complex $\omega-\bk$ plane, it was identified that a positive $\Im\omega$ is the true marker of instability, while a complex $\bk$ for some real $\omega$ is not. Important insight from Landau's treatment of two-stream instabilities in plasmas, allowed the dispersion relations to be developed into a comprehensive tool to study and categorize collective neutrino instabilities. In particular, it was understood that instabilities can be convective or absolute, depending on whether they drift as they grow, or grow in situ~\cite{Capozzi:2017gqd}. The understanding of pinching singularities and critical points~\cite{Capozzi:2017gqd, Yi:2019hrp}, following the analyses of Sturrock and Briggs developed in the context of plasmas, was shown to be useful for this purpose. This connection has been significantly developed in more recent work. In the context of homogeneous fast instabilities, the relevance of Penrose's principal value condition as well as the Nyquist criteria was noted~\cite{Fiorillo:2023a,Fiorillo:2023b,Fiorillo:2024e}. It was shown that in inhomogeneous case, the interaction energy is not conserved and draws from the reservoir of neutrino kinetic energies~\cite{Fiorillo:2024a}. Using a linear response approach, Landau damping was established via Case-van\,Kampen continuum modes~\cite{Fiorillo:2024b}. A key insight is that subluminal flavor waves can feed on resonant neutrinos, leading to weak instabilities~\cite{Fiorillo:2024c}. Slow oscillations have been analyzed, though primarily in the homogeneous and/or axisymmetric limit~\cite{Fiorillo:2024pns,Fiorillo:2025a}.

However, despite these important advances, most of the discussions have either been limited to fast oscillations or to special sub-regimes of slow oscillation. It was not clear how the previous body of results (esp. the Penrose conditions)  generalized to the physical case where the distribution function $g_\bG$ depends on both $E$ and $\bv$. After all, for an unstable system, qualitatively new instabilities can arise from small changes. For example, including the $\we=\Delta m^2/(2E)$ term allows $\omega=\bk\cdot\bv+\we$ to be satisfied by superluminal modes, and an instability can arise where none could have existed. Our work is a response to this gap in our current understanding. We have provided the tools to define, analyze, and understand the instabilities in the general case.

Though our goal was mainly to establish an existence result, a key application of our work will be to optimize searches for instabilities. As preempted in Ref.\,\cite{Dasgupta:2018ulw}, the so-called ``zero-mode'' is a conservative diagnostic of instability. Provided that there is a crossing where the principal value $\mI_\text{PV}<0$, i.e., Condition\,2 is satisfied for some $\bk$, choosing $k\to0$ makes it maximally negative in the fast limit. Two ways in which the $k=0$ mode fails to diagnose the instability is if the distribution has multiple crossings that violate Condition\,2b and evade winding, and\,/\,or the $\we$-dependent effects are such that only inhomogeneous modes can be stable (e.g., in case of time-like eigenvectors and inverted mass ordering, if one has $\partial g /\partial \we\ll0$ at all crossings; see Ref.\,\cite{Fiorillo:2024pns} for a contrasting example where the instability is for the ``wrong'' slope, because the eigenvectors are space-like).

It is also important to emphasize that our treatment, at the level of a linearized stability analysis, does not say anything about the fate of the system at late times. Based on analytical and numerical studies by several groups, it appears likely that flavor depolarization~\cite{Bhattacharyya:2020dhu,Bhattacharyya:2020jpj,Bhattacharyya:2022eed} is a key observable signature of fast collective effects~\cite{Wu:2021uvt,Richers:2022bkd,Nagakura:2022kic,Liu:2025tnf, George:2024zxz}; see also refs.~\cite{Nagakura:2019sig,Richers:2021nbx,Sigl:2021tmj,Richers:2021xtf,Abbar:2023zkm, Richers:2024zit, Padilla-Gay:2025tko}.  

Also, we have neglected effects due to momentum and/or number changing collisions. They are expected to be responsible for generating the flavor-dependent phase space distributions~\cite{Capozzi:2018clo,Martin:2021xyl}, but otherwise subdominant to fast oscillations. More recently, instabilities arising from collisions have been explored in refs.~\cite{Johns:2021qby,Lin:2022dek,Xiong:2022vsy,Padilla-Gay:2022wck,Johns:2022yqy,Xiong:2022zqz,Shalgar:2023aca,Akaho:2023brj,Zaizen:2025ptx, Froustey:2025nbi}. We defer these to future work.

\vspace{-0.3cm}

\subsection{Summary of Results}
\label{sec:summ}

\begin{enumerate}
\item In general, crossings of the distribution $g_\bG$  are not sufficient for collective instability. A ``positive slope'' and ``sufficiently negative principal value'' at the crossing are required, in addition, for the case of time-like eigenvectors and inverted ordering, for example. See Sec.\,\ref{sec:prop_suff_cond} for the precise proposition and Sec.\,\ref{sec:proof_suff_cond} for the proof. The remarks following the proof describe how to adapt the conditions for mass ordering and the causal character of the eigenvectors.

\item It is important to reiterate what our work does not do. It does not establish a one-to-one connection between the crossings of the distribution $g_\bG$ and the presence of instabilities. In fact, we showed that such a connection does not exist in general. The eigenvector, as well as more global properties of $g_\bG$, encoded via Conditions 2a and 2b, are also needed to determine whether instabilities are present. Our treatment precisely demonstrates where this connection fails and what additional information is needed. Though our approach provides guidance in identifying potentially unstable modes, as well as in ruling out instability in some cases, it does not provide an exhaustive method to identify all the unstable modes.

\item The core methods of this work are in Sec.\,\ref{sec:suffcon}. Key steps include: the scalar version of the dispersion relation in Eq.\,(\ref{eq:scDR}), the principal value integral in Eq.\,(\ref{eq:Ipv2}) and the Dirac delta integral in Eq.\,(\ref{eq:Id2}) for the multi-dimensional setting, the ansatz for shifted coordinates in Eq.\,(\ref{eq:coordguess}), the near-far separation in Eq.\,(\ref{eq:nearcon}), and the argument principle in Eq.\,(\ref{eq:AP}).

\item For fast oscillations, a crossing of $G_\bv$ is sufficient for instability. Due to the compactness of the space of velocities $\mathbb{S}^2$ and the closed nature of zero-crossing curves on it, the slope of the crossing as well as the inequalities of the principal value integrals required by the global conditions can be satisfied; see  Sec.\,\ref{sec:fastloose}. 

\item For slow oscillations, i.e., if the $\bv$-dependence of the distribution is negligible, crossings must have positive slope $\partial g /\partial \we >0$ to furnish an instability for time-like eigenvectors; see  Sec.\,\ref{sec:slowcross}. Further, in this case,  the smallest principal value $\mI_\text{PV}(\omega_0)$ must be sufficiently negative. There are also further global conditions on other principal value integrals. These global conditions may be satisfiable via approximate $\lambda$-scaling in the limit of small $\we$; see Sec.\,\ref{sec:slowloose}.

\item In the general mixed case, where the distribution has crossings along surfaces that vary with both $\bv$ and $\we$,  a more extensive set of conditions needs to be met, in addition to having a crossing. These depend on both local and global properties of $g_\bG$, the wave-vector $\bk$, and the eigenvector concerned. The global constraints are nontrivially dependent on the eigenvector through the factor of $|1-{\bf a}\cdot{\bf v}|^2$ in $g_\bG$. However, for large collective potential $\mu\gg \we$, the dependence typically softens via $\lambda$-scaling to the causal character alone. An unstable wave-vector $\bk$ needs to have a minimum length $k_\text{min}$\,(see  Sec.\,\ref{sec:mixcross}), depending on $\we$ and the gradients of the distribution, $\nabla_\bv g$  and $\partial g /\partial \we$, at the crossings. For $\partial g /\partial \we >0$, the only restriction comes from the extent of approximate $\lambda$-scaling; see Eq.\,(\ref{eq:welamlim}); for $\partial g /\partial \we <0$, one also has $k_\text{min}>0$ that represents an additional constraint on $\bk$; see  Sec.\,\ref{sec:mixloose}.

\item The significance of each of Conditions 1a, 1b, 2a, and 2b, as well as of $\lambda$-scaling, is shown through a set of numerical examples in Sec.\,\ref{sec:examples}.

\item The sufficiency and necessary conditions for instability have been explicitly proved and analyzed for time-like eigenvectors of the dispersion tensor.  Analogous conditions for space-like eigenvectors, with $1-|{\bf a}|^2<0$, and light-like (see Sec.\,\ref{sec:comments}) eigenvectors can be proved straightforwardly. These solutions are discrete poles of the dispersion relation.  However, a full classification of all solutions, including Landau damped and gauge-constraints, requires a treatment beyond mode analysis, by including the branch cuts. Our work is limited to identifying the conditions for instabilities.

\item Finally, our work has a close connection to a classic result by Penrose, on the stability of plasma oscillations. In a sense, it is a generalization of Penrose's conditions to ultra-relativistic particles with small but nonzero mass; see \,Sec.\ref{sec:penrose}.
\end{enumerate}

\section*{Acknowledgements} It is a pleasure to thank Damiano Fiorillo and Georg Raffelt for several helpful comments and discussions. We also thank Manuel Goimil-Garc\'ia and Lucas Johns for helpful remarks. This work is supported by the Dept.~of Atomic Energy~(Govt.~of~India) research project RTI 4002, and by the Dept.~of Science and Technology~(Govt.~of~India) through a Swarnajayanti Fellowship to BD.

\section*{Appendix: Proof of  Eq.\,(\ref{eq:IPVIDelta})}

We introduce a book-keeping parameter $s$ to explicitly keep track of orders in $\Delta m^2$, e.g., $\omega_i=\omega_i^{\rm fast} + s\omega_i^\Delta + {\cal O}(s^2)$. We will work to linear order in $s$ in the following.

To evaluate the integral, we choose a fixed direction for $\bk\,||\,\hat z$, without loss of generality, so that $\bk\cdot\bv=kv_z$. The momentum-space integral reduces to a two-dimensional integral over $x\equiv v_z$ and $y\equiv\we$,
\begin{equation}
    \mI_{\rm PV}(\omega_i) =\pv\int dy\,dx \,\frac{g(x,y)}{\omega_i - kx-sy}\,, 
\end{equation}
where we scale $y$ in the denominator to track the powers of $s$ and absorb the measure factor of $1/\we^4$ (and other constants) into a redefinition of the spectrum $g(x,y)$. We remember that we have assumed that the apparent singular nature of the integral at $\we=0$ will be cured by a suitable decay of $g_\bG$, and we will therefore take $g(x,y)$ to be analytic. At this stage, the pole is along a straight line in the $x-y$ plane. The variable transformation
\begin{align}
    x^{\prime} &= \;x - \frac{s(\omega_i^\Delta-y)}{k}\equiv x-\sigma(y)\,,\\
    y^{\prime} &= \; y\,,
\end{align}
leads to poles at $x^{\prime}=\omega_i^{\rm fast}/k\equiv x_0$ and the Jacobian being 1. Replacing the dummy variables $x'$ and $y'$ by $x$ and $y$, respectively, we have
\begin{align}
    \mI_{\rm PV}(\omega_i) &=\int dy \,\pv\int_{-1-\sigma(y)}^{1-\sigma(y)} dx \,\frac{g(x + \sigma(y),y)}{k(x_0-x)}\,,
\end{align}
where the $x$-argument of $g$ and the $x$-integration range are now shifted by $\pm \sigma(y) \propto s$, respectively. At fixed $y$, Taylor expanding the function $g\left(x + \sigma,y\right)$ about $x$ to linear order, we get
\begin{align}
    \mI_{\rm PV}(\omega) &=\int dy\,\pv \int^{1-\sigma(y)}_{-1-\sigma(y)} dx \,\frac{g(x,y)}{k(x_0-x)} \,\nonumber\\
    &+\,  \int dy\, \sigma(y)\,\pv\int^{1-\sigma(y)}_{-1-\sigma(y)} dx \,\frac{\partial_{x}g(x,y)}{k(x_0-x)}\,.
\label{eq:IPVmix}
\end{align}
Further, for $f(x)=g(x)/(x_0-x)$ or $\partial_x g(x)/(x_0-x)$ with $g(x)$ being analytic and $f(x)$ independent of $\sigma$, to linear order we have
\begin{align}
\int^{1-\sigma}_{-1-\sigma} dx \,f(x) &= \int^{1}_{-1} dx \,f(x) +  \sigma\frac{d}{d\sigma}\Bigg\vert_{\sigma=0}\int^{1-\sigma}_{-1-\sigma} dx \,f(x)\,\\
&=\int^{1}_{-1} dx \,f(x) + \sigma \Big[-f(1) + f(-1) \Big]\,,\label{eq:strick}
\end{align}
where we used Leibniz rule in the last step. Applying this to Eq.\,(\ref{eq:IPVmix}), the range of the $x$ integration reverts to $(-1,1)$ for both the first term (as the spectrum $g(x,y)$ vanishes at $x =\pm1$ when neutrinos have nonzero mass), and the second term (as it is an order-$s$ term, and the difference is order-$s^2$). 

\pagebreak
Finally, setting the book-keeping parameter $s$ to unity, we have that which we set out to prove:
\begin{align}
    \mI_{\rm PV}(\omega_i) &=\int d\we\,\pv \int^{1}_{-1} dv_z \,\frac{g(v_z,\we)}{\omega^{\rm fast}_i - kv_z} \,\nonumber\\
    &+\,  \int d\we\, \frac{\omega^\Delta_i-\we}{k}\,\pv\int^{1}_{-1} dv_z \,\frac{\partial_{v_z}g(v_z,\we)}{(\omega^{\rm fast}_i - kv_z)}\,,\label{eq:Ipvrobust}
\end{align}
where we can identify the first term with $\mI_{\rm PV}^{\rm fast}$ and the second term as the correction $\mI_{\rm PV}^{\Delta}$. To obtain $\mI_{\rm PV}^{\Delta}$ in the form displayed in Eq.\,(\ref{eq:IPVIDelta}), we can simply integrate by parts to get
\begin{equation}
    \mI_{\rm PV}^{\Delta} = \pv\int d\bG \,\frac{\big(\we-\omega_i^\Delta\big)\,g_\bG}{(\omega^{\rm fast}_i - \bk\cdot\bv)^2}\,,
\end{equation}
where we have again dropped the (vanishing) boundary terms at $v_z=\pm 1$. The second term now looks more singular, and the double-pole would usually not allow a principal value treatment. However, because of the vanishing of $g_\bG$ at the boundary, it is well-defined in a principal value sense; this is clear from Eq.(\ref{eq:Ipvrobust}) which only has a simple pole.

It is not essential to assume that the spectrum vanishes at $v_z=\pm1$, as we have done here. More generally, one has the boundary terms (cf. Eq.\,(\ref{eq:strick})), but they too scale as $1/\lambda^2$.

\linespread{0.93}

\providecommand{\href}[2]{#2}\begingroup\raggedright\endgroup

\end{document}